\documentclass[
 reprint,
 amsmath,
 pre,
 amssymb,
 nofootinbib,
 aps,
 longbibliography, 
superscriptaddress
]{revtex4-1} 

\usepackage[pdftex]{graphicx}
\usepackage{wasysym}
\usepackage{amsfonts}
\usepackage{ifsym}
\usepackage{pifont}
 \usepackage{footmisc}
\usepackage[export]{adjustbox}
\usepackage{dsfont}
\usepackage{lipsum}
\usepackage{xr}
\usepackage{subfiles}
\usepackage{braket}

\makeatletter
\newlength{\apb@width}
\newcommand{\autoparbox}[2][c]{\settowidth{\apb@width}{#2}\parbox[#1]{\apb@width}{#2}}

\makeatother

\newcommand{\namedref}[2]{\hyperref[#2]{#1~\ref*{#2}}}

\renewcommand{\Re}{\mathop{\mathrm{Re}}}
\renewcommand{\Im}{\mathop{\mathrm{Im}}}

\newcommand{\Csphere}{{}^\bullet\kern-1.2pt C}
\newcommand{\Ctorus}{{}^\circ\kern-1.2pt C}

\newcommand{\nn}{\nonumber}

\newcommand{\COMMENT}[1]{}

\newcommand{\neqa}{\nonumber\end{eqnarray}}
\newcommand{\la}[1]{\label{#1}}

\newcommand{\<}{{\langle}}
\renewcommand{\>}{{\rangle}}

\newcommand{\re}{\relax{\rm I\kern-.18em R}}

\def\su2{{SU(2)}}

\def\[{\left[}
\def\]{\right]}

\def\({\left(}
\def\){\right)}
\def\[{\left[}
\def\]{\right]}

\def\<{\langle}
\def\>{\rangle}

\def\i2{\frac{i}{2}}

\def\2F1{\,_2{\rm F}_1}

\usepackage[usenames,dvipsnames]{xcolor}
\usepackage{color}
\usepackage{graphicx}
\usepackage{dcolumn}
\usepackage{bm}
\usepackage[hidelinks]{hyperref}

\usepackage{cancel}
\usepackage{ctable}
\usepackage{booktabs}
\newcolumntype{L}[1]{>{\raggedright\let\newline\\\arraybackslash\hspace{0pt}}m{#1}}
\newcolumntype{C}[1]{>{\centering\let\newline\\\arraybackslash\hspace{0pt}}m{#1}}
\newcolumntype{R}[1]{>{\raggedleft\let\newline\\\arraybackslash\hspace{0pt}}m{#1}}

\newcommand{\beq}{\begin{equation}}
\newcommand{\eeq}{\end{equation}}
\newcommand{\beqq}{\begin{equation*}}
\newcommand{\eeqq}{\end{equation*}}
\newcommand\beqa{\begin{eqnarray}}
\newcommand\eeqa{\end{eqnarray}}
\newcommand\beqaa{\begin{eqnarray*}}
\newcommand\eeqaa{\end{eqnarray*}}
\newcommand\bea{\begin{array}}
\newcommand\eea{\end{array}}

\usepackage{MorrisIn,lettrine,Romantik}

\usepackage[toc,page,titletoc]{appendix}
\usepackage[capitalise,nameinlink]{cleveref}

\crefname{supp}{Supplement}{Supplements}

\begin{document}

\title{Multiparticle Flux Tube S-matrix Bootstrap}
\author{Andrea Guerrieri}
\affiliation{Perimeter Institute for Theoretical Physics, 31 Caroline St N Waterloo, Ontario N2L 2Y5, Canada}
\affiliation{Dipartimento di Fisica e Astronomia, Universita degli Studi di Padova, Italy \\ \& Istituto Nazionale di Fisica Nucleare, Sezione di Padova, via Marzolo 8, 35131 Padova, Italy}
\author{Alexandre Homrich}
\affiliation{Perimeter Institute for Theoretical Physics, 31 Caroline St N Waterloo, Ontario N2L 2Y5, Canada}
\author{Pedro Vieira}
\affiliation{Perimeter Institute for Theoretical Physics, 31 Caroline St N Waterloo, Ontario N2L 2Y5, Canada}
\affiliation{Instituto de F\'isica Te\'orica, UNESP, ICTP South American Institute for Fundamental Research, Rua Dr Bento Teobaldo Ferraz 271, 01140-070, S\~ao Paulo, Brazil}

\begin{abstract}

We introduce the notion of branon jets, states of collinear flux tube excitations. We argue for the analyticity, crossing and unitarity of the multi-particle scattering of these jets and, through the S-matrix bootstrap, place bounds on a set of finite energy multi-particle sum rules. Such bounds define a matrioska of sorts with a smaller and smaller allowed regions as we impose more constraints.
The Yang-Mills flux tube, as well as other interesting flux tube theories recently studied through lattice simulations, lie inside a tiny island hundreds of times smaller than the most general space of allowed two-dimensional theories.

\end{abstract}

\pacs{Valid PACS appear here}
\maketitle

\section{Introduction}

  \lettrine{S}{catter} at high energies and particles you shall produce. Such is life in a relativistic world; $2\to n$ amplitudes exist. And yet, we are often afraid of them, and rightfully so. Even the number of variables required to describe their kinematics is intimidating, not to speak of their intricate analytic structure, and it is perhaps not surprising that within the recent resurgence of the S-matrix bootstrap 
  only $2\to 2$ amplitudes have been constrained thus far. In this paper, we give a first step towards improving this state of affairs and delve into the $m \to n$ world. We will do so within the physics of long one-dimensional flux tubes.  

The transverse fluctuations of long flux tubes are described by massless particles, Goldstone bosons of the nonlinearly realized Poincar\'e symmetry \cite{Isham:1971dv,Volkov:1973vd,Dubovsky:2012sh,Aharony:2013ipa}.
We call them branons. The flux tube S-matrix bootstrap \cite{FluxTube,EliasMiro:2021nul,Gaikwad:2023hof} aims at studying these flux tubes by constraining the possible S-matrices of these branons. We will focus on flux tubes in three space-time dimensions so that we have a single goldstone particle.

Its two-to-two scattering at low energies is given by\footnote{For simplicity we will sometimes refer to $s$ -- the center of mass energy \textit{squared} measured in string length units -- as \textit{the energy}. The $768$ factor in (\ref{Sexp}) leads to good $O(1)$ numbers for $\gamma$, see e.g.~(\ref{gamma}) and lattice estimates reviewed below}  
\beq
S_{11\to 11} (s) = e^{i s \ell^2/4} \times e^{i \gamma \ell^6 s^3/768} \times (1+O(s^4))  \la{Sexp}
\eeq
where $\ell$ is the string length (set to one henceforth) and where we separated the amplitude into three parts: (a) a "gravitational" dressing \cite{Conkey:2016qju,Dubovsky:2017cnj,dressing} which would arise from the Nambu-Goto string action, (b) the first deviation from NG parametrized by the Wilson coefficient $\gamma$ and (c) all higher corrections which UV complete this amplitude. The existence of such UV completion with $|S|\le 1$ for all positive energies together with the assumption of polynomial boundness in the upper half plane leads to a non-trivial bound on the first Wilson coefficient as \cite{FluxTube}
\beq
\gamma \ge - 1 \,. \label{gamma}
\eeq
Here we suggest the use of multi-branon processes to further constrain the dynamics of quantum flux tubes.\footnote{While multi-particle processes will highly constrain the dynamics of flux tubes at intermediate energies, leading low energy constraints such as (\ref{gamma}) will \textit{not} be improved by including them, see also appendix \ref{toy_production}. \label{gammabound}} 
We will produce the first S-matrix bootstrap bounds involving multi-particle processes.

The key player in our construction is what we denote as a \textit{branon jet}, a multi-particle state with $n$ particles moving collinearly, each with a fraction $\alpha_i$ of the total energy $P$
\beq
|\alpha_1,\dots,\alpha_n, P\>_n \equiv |\alpha_1 \vec{P},\alpha_2 \vec{P} ,\dots, \alpha_n \vec{P} \> 
\label{branon_jet_def}
\eeq
where $\alpha_1+\dots+\alpha_n=1$. 

The jet can be left-mover or right-mover and within a (left) right-mover jet all constituents are (left) right-movers. 
It is important to stress that the possibility of making sense of such a state of collinear massless particles is quite non-trivial. 
It is well-defined due to the absence of collinear divergences for these goldstone particles \cite{Dubovsky:2012wk}. In fact, the scattering of any number of left movers (or right movers) amongst themselves is trivial, 
\beq
S_{RR'}=S_{LL'}=1 \label{LLRR}
\eeq
since we can boost such processes to arbitrarily low energy when they become effectively free. This is true for $R$ and $R'$ being jets of any number of right movers or simply fundamental right-moving individual branons. 
These crucial properties allows us to cross jets of multiple particles from past to future, argue for the all-loops analyticity of their scattering amplitudes, and treat them effectively as new massless fundamental particles; their internal composition can be simply thought as a new flavor index, see appendices \ref{theoryappendix} to \ref{higherjetsection}. 

For jets of two particles (and similarly for jets of many particles) we define a discrete complete basis  by averaging over the energy fraction $\alpha$ at fixed total energy $P$ according to
\beq
|n,P\> \equiv \! \sqrt{2n+1} \! \int\limits_0^1 \! d\alpha \frac{P_n(2\alpha-1)}{\sqrt{8\pi \alpha(1-\alpha)}}   |\alpha,(1-\alpha), P\>_2  \, \label{jet2}
\eeq
with $n$ even since branons are identical bosons, see appendix \ref{higherjetsection} for the general treatment of $N$ particle jets. 
The precise normalization and choice of Legendre polynomials for jet wave functions in (\ref{jet2}) is  tuned so that these jets are nicely normalized as one-particle states 
\beq
\< m, Q |n,P\> = 4\pi P_0 \delta_{n,m} \delta(\vec{Q}-\vec{P}) \,. \nonumber
\eeq

We can now decompose a multi-particle amplitude into infinitely many scattering amplitudes involving these jets! We can scatter one jet against a fundamental branon, for instance, to get two possible S-matrices, $S_{1   n \to 1   n}(s)$  and $S_{1   n \to   n 1 }(s)  $ 
corresponding to the forward and backward scattering of the jet.

At low energies, all these S-matrices,
\begin{eqnarray}
&&\hspace{0.1cm}\!\!\!\!\!\!\!\!S_{11 \to 11} =\! \includegraphics[height=0.8cm,valign=c]{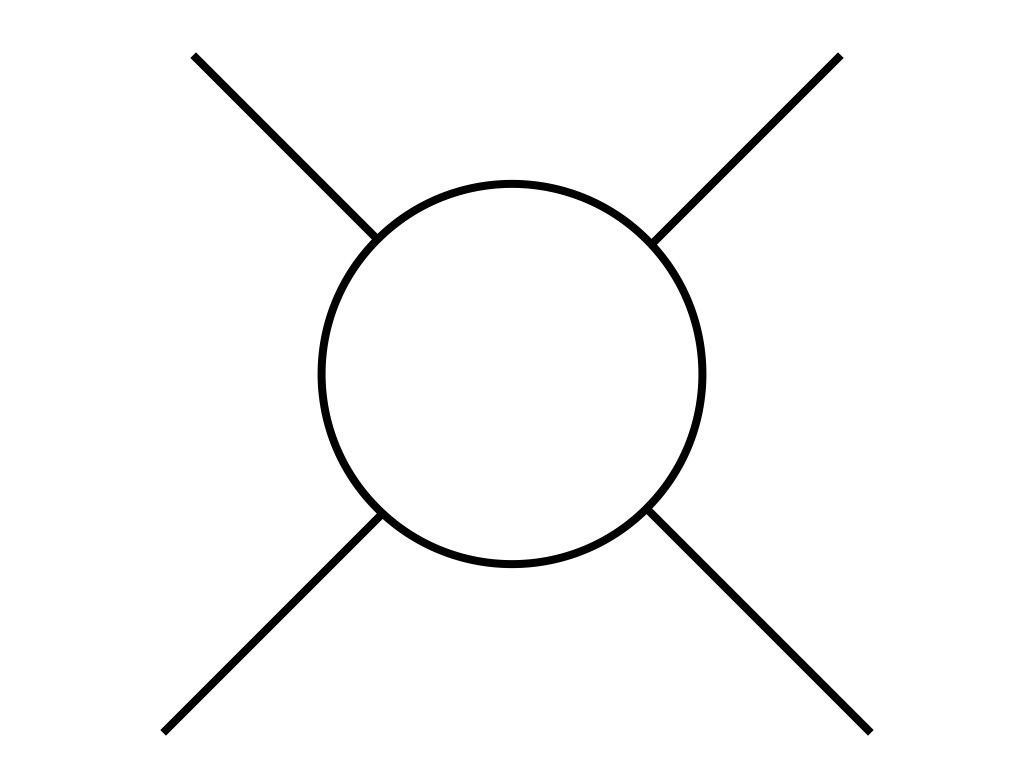} ,  S_{1{  n} \to 1 {  m}  } = \! \includegraphics[height=0.8cm,valign=c]{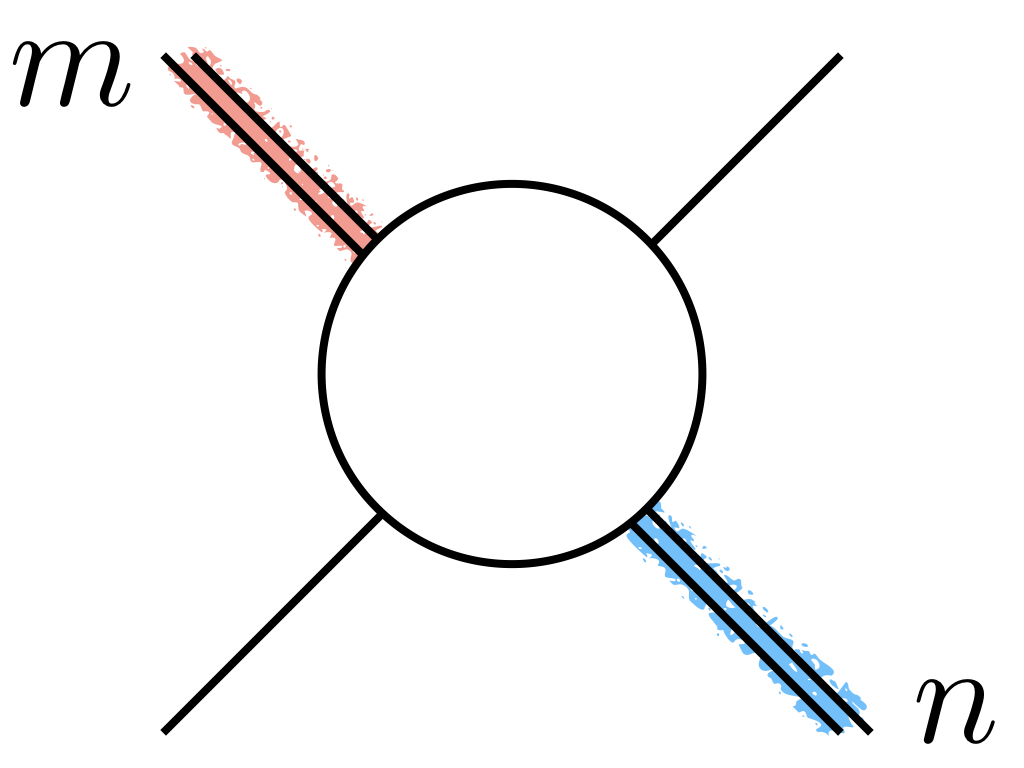} \,,  S_{{  n}1 \to {  m} 1 } = \! \includegraphics[height=0.8cm,valign=c]{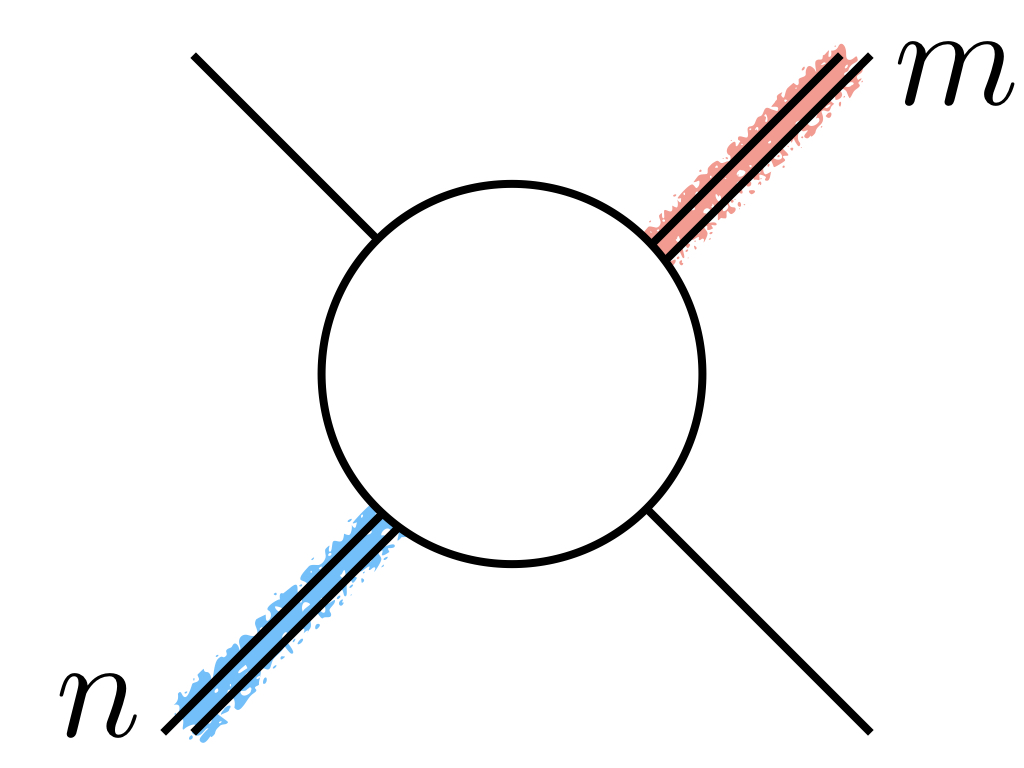} \,,\nn \\
&&\hspace{0.1cm}\!\!\!\!\!\!\!\!S_{ {  n} 1 \to 1 {  m} } = \! \includegraphics[height=0.8cm,valign=c]{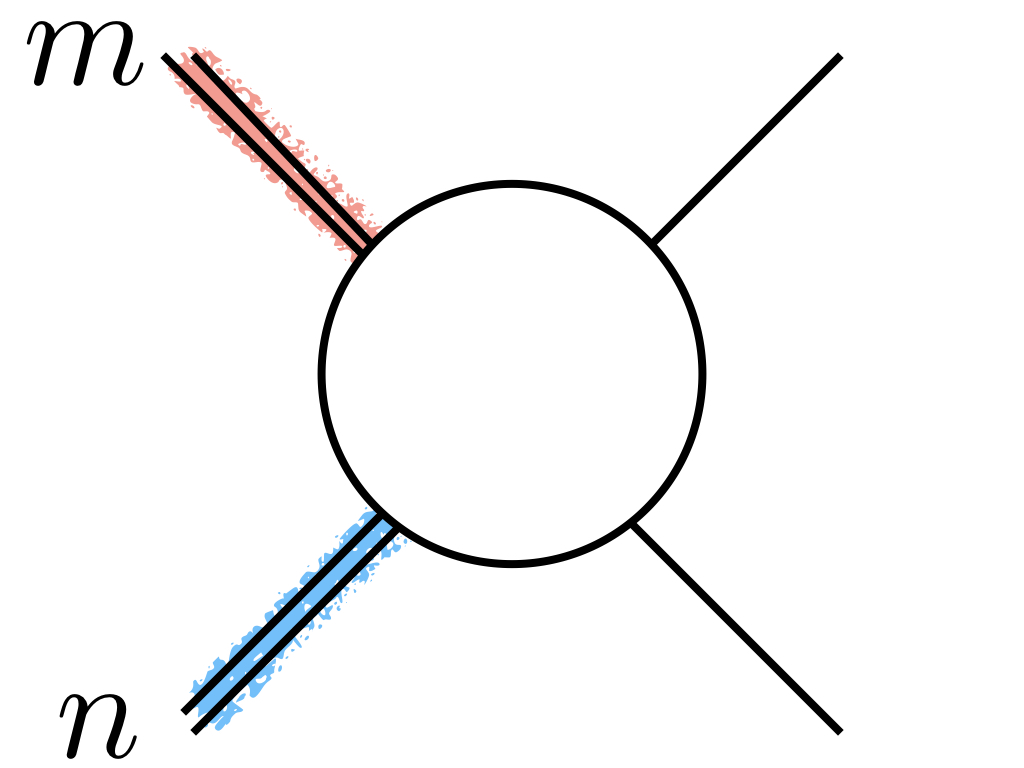} \,, S_{ 1{  n}  \to  {  m} 1} =\!  \includegraphics[height=0.8cm,valign=c]{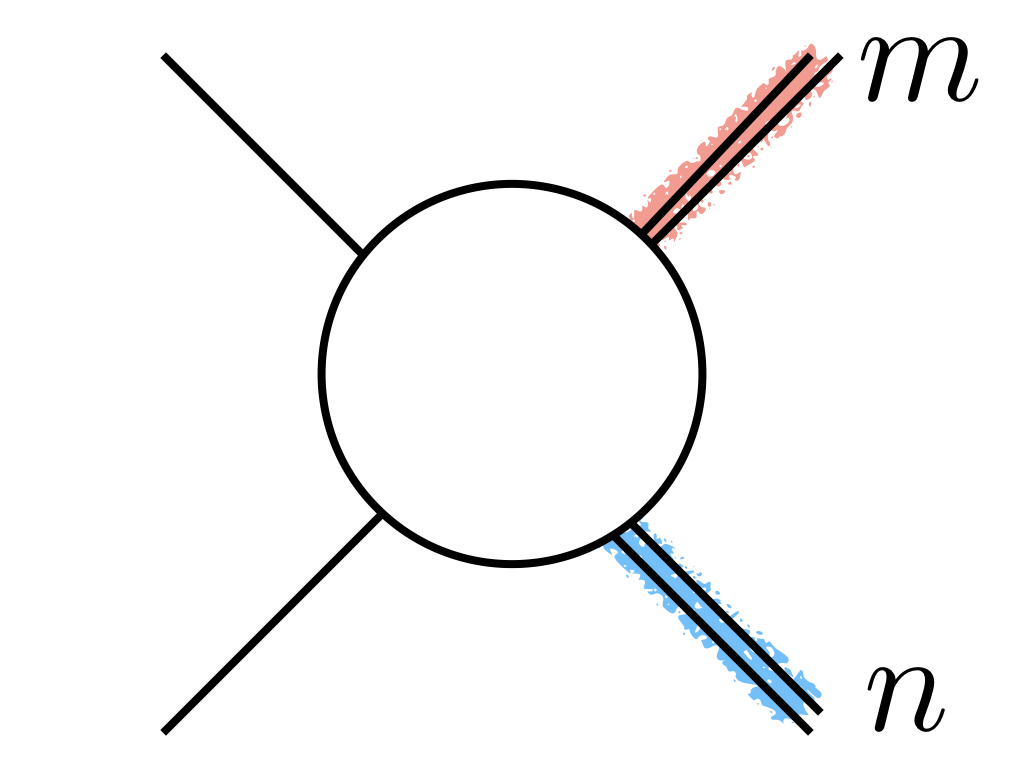} \,, S_{11 \to {  n} {  m} } =\!\!  \includegraphics[height=0.8cm,valign=c]{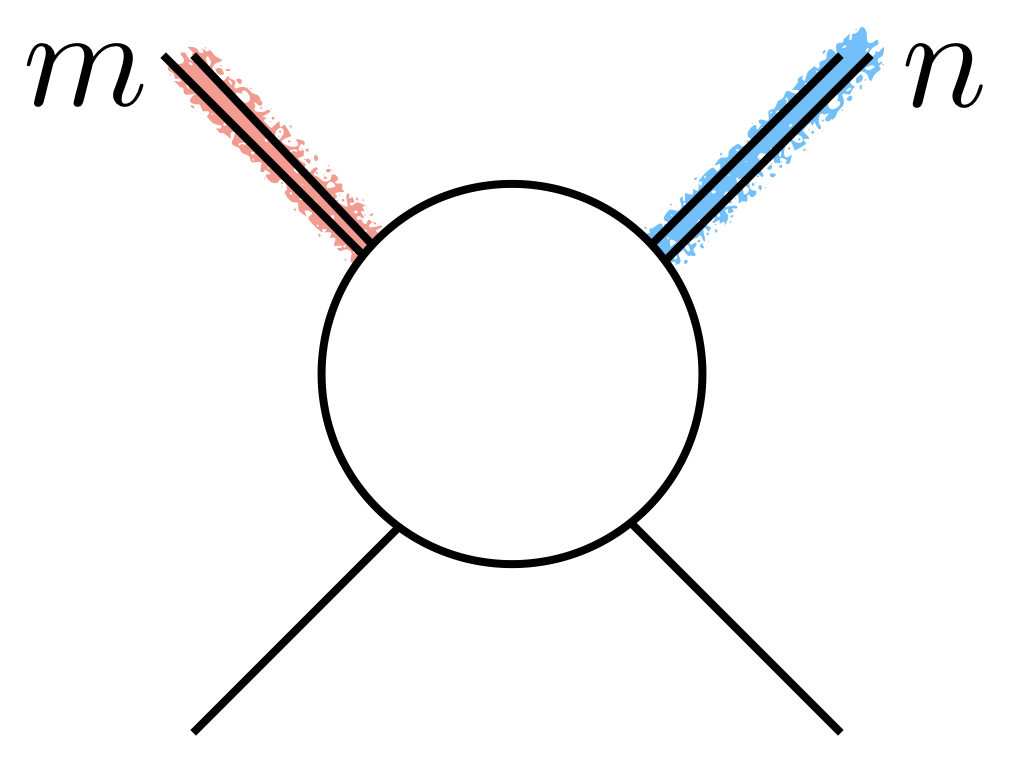} \,, \nn \\
&&\hspace{0.1cm}\!\!\!\!\!\!\!\!S_{ {  n} {  m} \to 11} =\!  \includegraphics[height=0.8cm,valign=c]{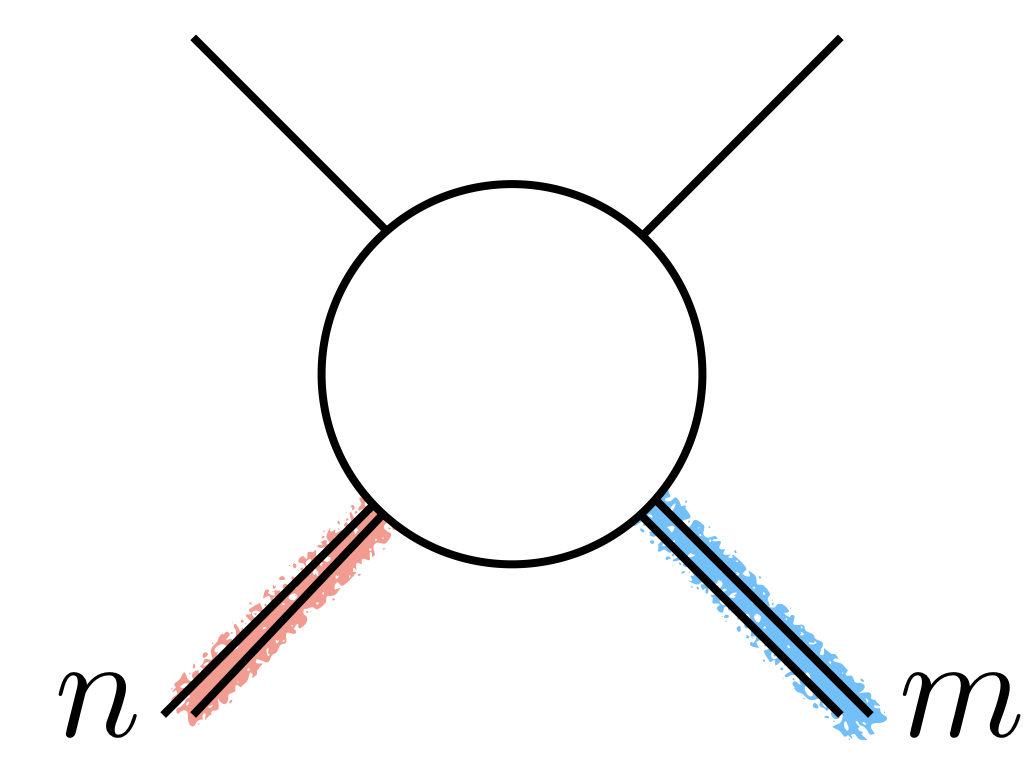} \text{ and finally } S_{{  p} {  n} \to {  r} { m} }= \! \includegraphics[height=0.8cm,valign=c]{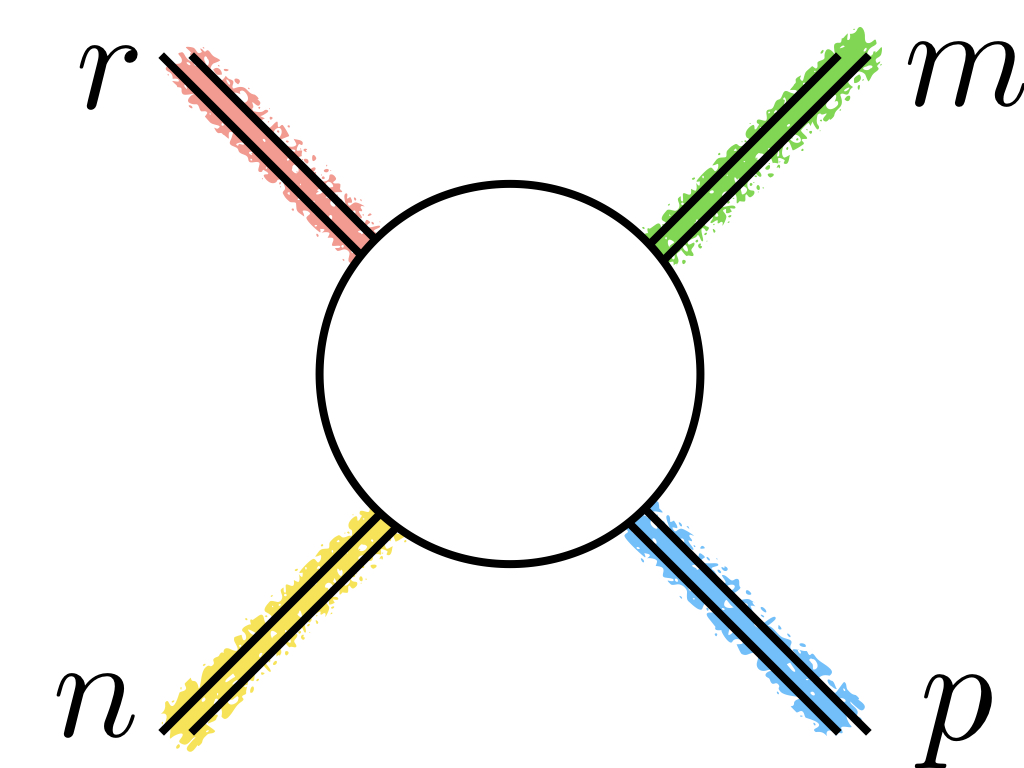} \,, \nn
\end{eqnarray}
are again given by a ``gravitational" dressing part $e^{i {\ell^2} P_L P_R/4} $
where $P_{L}/P_R$ is the total left/right momenta of the incoming (or outgoing) state together with the first deviation from this Nambu-Goto behaviour, governed by $\gamma$ and captured concisely by the generating function \cite{dressing} 
\beqa
\!\! L_\gamma= 
\frac{\gamma}{3\cdot 2^{13}}
(\partial_+^2 x)^2 (\partial_-^2 x)^2  \big( 16   \ell^6+24 \ell^8  (\partial_+ x) (\partial_- x)+ \,\, \nn \\
\qquad \qquad\qquad +21  \ell^{10}   (\partial_+ x)^2 (\partial_- x)^2+\dots \big), \label{lagrangian}
\eeqa
 where $x$ in this equation indicates the transverse coordinate to the flux tube.
Combining both factors we can read off the low energy behavior of all scattering processes involving two-particle jets and fundamental particles as detailed in appendix \ref{apFD}.

The behaviour of all these S-matrices at energies $s \sim \ell^{-2}$ is non trivial. For most flux tube theories of interest, such as pure Yang-Mills, their dynamics is expected to be strongly coupled at those energies. In this short note we will place bounds on the simplest possible multi-particle observables capturing the dynamics at the string scale. For that we will consider only the fundamental branon -- for which we use the label $1$ -- together with the first branon jet given by $n=0$ in (\ref{jet2}) -- which we will henceforth indicate by the label $2$.

This nomenclature where we now have two effective particles with labels $1$ and $2$ is quite convenient. Indeed, note that we now have a simple S-matrix bootstrap setup with two particles and $\mathbb{Z}_2$ symmetry (with $1$ being the odd particle and $2$ the even one) and four independent amplitudes for which a great deal of technology has been developed in  \cite{Homrich:2019cbt,Bercini:2019vme,dual2}.

\section{A Triplet of Finite Energy Observables} \label{tripletS}
\lettrine{A}{ny} S-matrix bootstrap study starts with the choice of wise observables to bootstrap. We suggest\footnote{Note that real analyticity implies that $X$ and $Y$ are real. Also, $Z=\text{Re}(S_{11\to 22}(i))$ or $Z=\text{Re}(e^{i \phi} S_{11\to 22}(i))$ would all lead to exact same bounds since we can always rotate the in and out states $11$ and $22$ by a different phase (which is canceled out when doing processes with the same in and out particle types) so that there is no loss of generality in picking $\phi=0$ and defining our triplet as in (\ref{XYZ}). 
}
\beq
(X,Y,Z) \equiv (S_{11\to 11}(i),S_{22\to 22}(i),\text{Re}(S_{11\to 22}(i)))  \label{XYZ}
\eeq
as an interesting triplet of physical observables, a  finite energy section of the infinite-dimensional space of two-dimensional S-matrices.

The amplitudes $S_{AB \to CD}(s)$ are analytic in the upper half plane, so a natural observable measuring their strength can be defined as the value of these amplitudes somewhere in this upper plane; here we pick an energy~$s=i$. 

\begin{figure*}[t]
  \includegraphics[width=\textwidth]{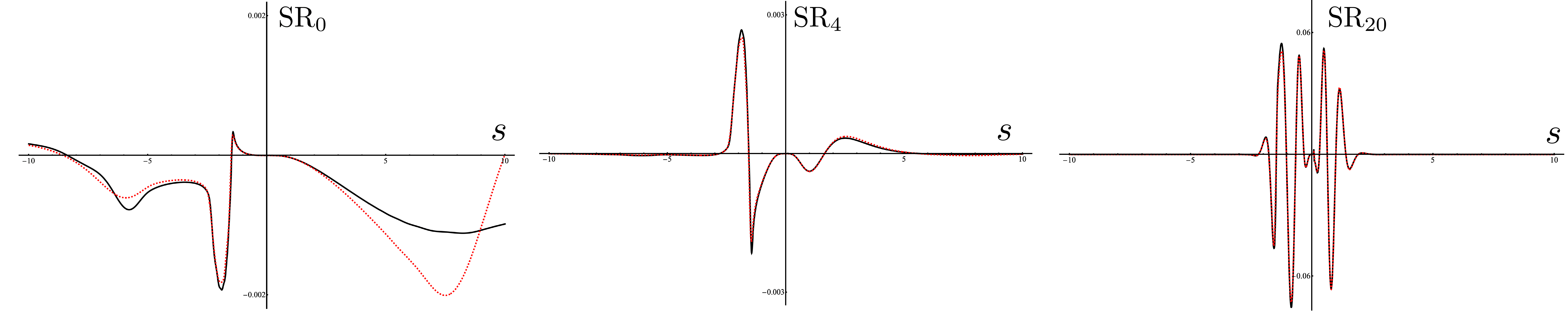}
  \caption{Different sum rules for different $n$'s lead to dramatically different integrands which nonetheless all integrate to exactly the same thing. Here we see that a low $n$ sum rule (on the left) decays slowly at large energies and is thus very sensitive to scattering there while a large $n$ sum rule (on the right) strongly suppresses high energy but requires a huge resolution at low energies which can also be a challenge. In the middle we depict a nice choice of intermediate $n=4$ which nicely suppresses large energy while not demanding a crazy resolution at low energy. In these plots the black and red curves represent two different Boostrap runs with two different $n_\text{max}$'s (130 and 75 respectively); we can think of them as two experiments with different resolution.} \label{SRn}
\end{figure*} 

What is $s=i$? \textit{If} we are dealing with flux tubes, the scale is set by the string length
so that $s=i$ means one imaginary unit measured in string length units. \textit{If}, on the other hand, we consider the most general space of  theories of massless particles with trivial collinear scattering  without any further low energy constraint to set a scale, then $s=i$ can be \textit{any} energy. If we establish that $X<0.7$, say, for some choice of $Y=Y_0$ and $Z=Z_0$ that means that $S_{11\to 11}$ is always smaller than $0.7$ for \textit{any} value of purely imaginary $s$ if the other components take the values $Y_0$ and $Z_0$ at \textit{that} energy. 

Let us also note that $(X,Y,Z)$ can be measured from simple sum rules probing scattering at all energies.
For example, we can write
\beq
X = \int\limits_{\mathbb{R}}\frac{ds}{\pi(s^2+1)} S_{11\to 11}(s)  \la{Xsum}
\eeq
and similarly for $Y$ and $Z$.\footnote{We could also further simplify this integral noting that $s>0$ and $s<0$ are simply related by crossing leading to $X=  \int_{0}^\infty \frac{2ds}{\pi(s^2+1)} \text{Im}(S_{11\to 11}(s)) $. For $Z$ the integral for $s>0$ and $s<0$ are not trivially related and we would get an integral over positive $s$ of the $S_{11\to 22}$ amplitude plus another integral over the same positive $s$ domain involving the (conjugate of the) $S_{12 \to 21}$ amplitude.} 

A flux tube experimentalist might like this sum rule: from several scattering outcomes at various energies $s$ one simply adds them up to produce a nice approximation to~(\ref{Xsum}). Of course, depending on the quality and quantity of such scattering data for different energy ranges, such experimentalist could also prefer other sum rules. Indeed, in~(\ref{Xsum}) we considered a single subtraction $s_0$ which is always enough for one dimensions; we could write infinitely many equivalent sum rules for $X, Y$ or $Z$ with more subtractions such as 
\beq
Z = SR_n \equiv \Re \int\limits_{\mathbb{R}}\frac{ds}{\pi(s^2+1)} \left(\frac{3i}{s+2i}\right)^n S_{11\to 22}(s)  \la{Xsumn}
\eeq
All these integrals must evaluate to the same thing although 
they weigh the real $s$ regions very differently. For large $n$ we suppress high energy but the low energy contribution becomes highly enhanced while for small $n$ one needs good high energy data but the low energy data needs not be resolved so much. An intermediate $n$ could be ideal for an experimentalist with reasonable but not ideal low and high energy data. Our bootstrap numerics below are a sort of gedankenexperiment and we illustrate this interesting $n$ region interplay in figure \ref{SRn}.  

To conclude: our triplet choice $(X,Y,Z)$ can be thought as a measure of the various contact multi-point couplings of the theory at some non-perturbative finite (complex) energy as well as a set of sum rules probing scattering across all (real) energy scales. We now turn to the allowed $(X,Y,Z)$ space.

\section{The Branon Matrioska} \label{MatrioskaS}
\lettrine{W}{hat} values can $(X,Y,Z)$ take? The answer depends on what physical conditions we impose. If we impose very little we will get that $(X,Y,Z)$ must leave inside a big (albeit compact) three-dimensional \textit{rock}. If we impose more conditions we will obtain a smaller \textit{rock} inside that one. And so on. We call this sequence of allowed regions -- one inside the other -- the \textit{Branon Matrioska}. We will now describe a Matrioska with three such regions.

\begin{figure*}[t]  \includegraphics[width=\textwidth]{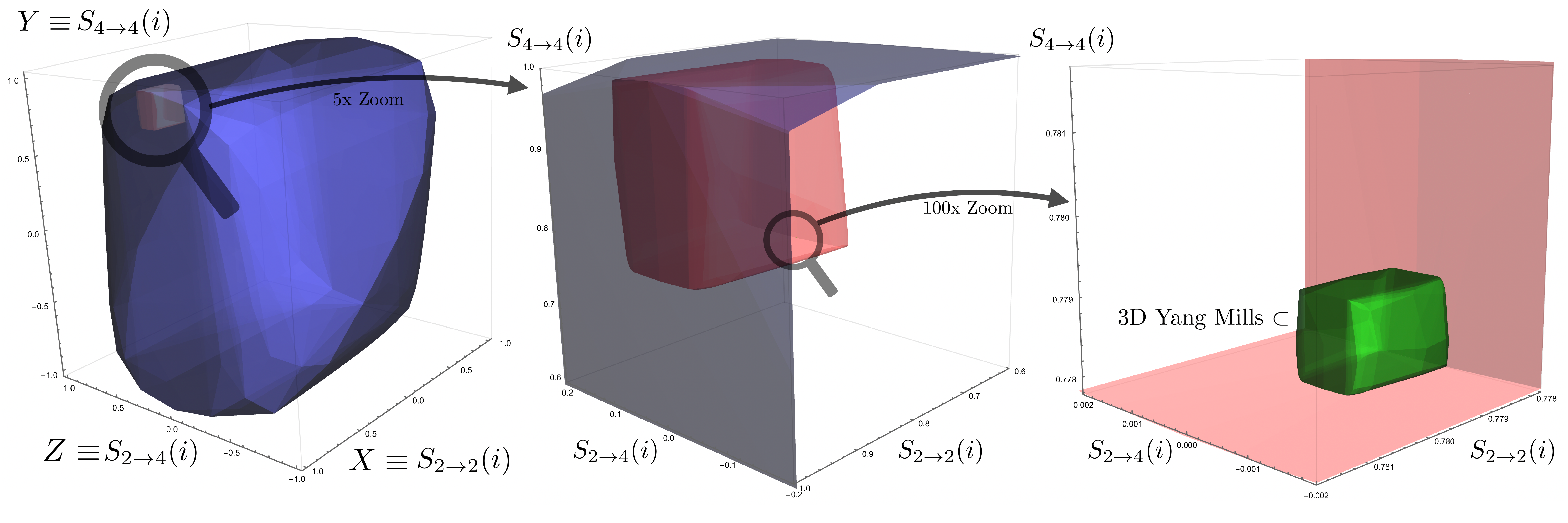}
  \caption{The Branon Matrioska: allowed space of $(X,Y,Z)$ for various two dimensional theories where jets make sense. The blue shape assumes only unitarity and analyticity. Inside it is a much smaller red surface where we also impose non-linearly realized Lorentz by fixing the low energy behavior of the S-matrices. Inside it, in a yet much smaller green region is the space of such S-matrices with first Wilson coefficient $\gamma \lesssim 0.8$, a conservative upper bound which should contain most of the interesting flux tube theories according to the recent lattice estimates. In this figure we use $S_{n\to m}$ to denote the processes involving in total $n$ and $m$ particles in the initial and final states; $S_{4\to 4}$ for instance refers to the scattering of two jets (each with two particles) in the past yielding two jets in the future, that is $S_{4\to 4}=S_{22\to 22}=Y$ and so on. In each matrioska doll, some directions can be bounded from analytic single-component Schwarz type arguments, see appendix \ref{analyticalbounds}.}. \label{MatrioskaFig}
\end{figure*}

The first physical constraint we will impose is unitarity. Unitarity can be stated as positive semi-definiteness of two simple matrices. One is obtained by constructing all possible scalar products of even in and out states so that 
\beq
\left( 
\begin{array}{cccc}
1 & S_{11\to 11} & 0 & S_{11\to 22} \\
S_{11\to 11}^* & 1 & S_{22\to 11}^* & 0 \\
0 & S_{22\to 11} & 1 & S_{22\to 22}\\
S_{11\to 22} ^* & 0 & S_{22\to 22}^* & 1
\end{array}
\right) \succeq 0 \label{unit}
\eeq
for any $s>0$. The other condition, for the odd sector, takes a similar form with $11$ and $22$ replaced by $12$ and $21$ respectively. 
These two unitarity conditions talk to each other through crossing \cite{Homrich:2019cbt} since $S_{11\to 22}= S_{12\to 21}^*$. It is this interplay between crossing and unitarity which leads to non-trivial bounds.

The positivity conditions \eqref{unit} are imposed numerically using \texttt{SDPB} \cite{Simmons-Duffin:2015qma}. In the primal formulation 
we explore a (truncated version of an) ansatz \cite{FluxTube}
\beq
S_{AB \to CD} = \sum_{m=0}^\infty c_m^{(AB\to CD)} \left(\frac{s-i}{s+i}\right)^m
\eeq
subject to unitarity and optimize over the constants~$c_m^{(AB\to CD)} $ to explore the allowed $(X,Y,Z)$ space. The truncation amounts to replacing $\infty$ by a large $n_\text{max}$ cut-off. For a dual formulation see appendix \ref{dualBlue} and \ref{DualRed}.  

For the outer doll of the Matrioska, this is all we impose: \textit{unitarity} and \textit{crossing} symmetry (plus \textit{analyticity} implicitly). Importantly, we do \textit{not} impose any low energy behavior. We call this problem the \textit{blue} problem as we will present its results in a blue $3D$ plot. Note right away that a particular obvious consequence of unitarity is that $|X|<1$, $|Y|<1$ and $|Z|<1$ since these are all probabilities and as such the blue rock must be a compact shape inside a cube of length $2$ centered at the origin of the $(X,Y,Z)$ space; indeed, this is what we find numerically, see blue shape in figure \ref{MatrioskaFig}. 

The \textit{red} problem is the setup where we impose unitarity together with the leading universal EFT
\begin{eqnarray}
S_{ab\to ab}=e^{is/4}+O(s^2) \,,\, S_\text{other components}=O(s^2) \label{EFTNG}
\end{eqnarray}
imposing dominance of the elastic components over the reflection and creation amplitudes at low energies. 
This is the universal behavior predicted by the leading EFT given by Nambu-Goto and is agnostic about the first correction to it governed by $\gamma$. Imposing a low-energy behavior can dramatically improve our finite energy bounds as reviewed in a few simple analytic examples in appendix \ref{goodP}. Indeed, it does! We immediately obtain a red rock about ten times smaller than the blue rock once we impose these extra constraints, see middle panel in figure~\ref{MatrioskaFig}.

As explained in  appendix \ref{apFD}, we know the behavior of all flux tube S-matrices a few orders further in the low energy expansion up to the terms governed by the leading Wilson correction $\gamma$. 
The reader could wonder 
whether imposing that subleading behavior 
would generate another rock inside the red rock.
We have indeed tried that but all numerical evidence we found is that it does \textit{not} improve the bounds further; the resulting rock seems to converge to the very same red rock as we increase our primal truncation $n_\text{max}$. This is fine: not all pointwise constrains lead to bound improvements as explained in the toy examples of appendix \ref{goodP}. This seems to be one more such example -- as long as we do not commit to any value for $\gamma$. 

On the other hand, if we also supplement our conditions with an upper bound on the Wilson coefficient $\gamma$ the situation dramatically improves. This is what we did in what we call the \textit{green} problem. Here we imposed the constraint
$\gamma<0.768$ which we expect should include all non-abelian pure glue gauge theories according to various recent lattice estimates\footnote{For $SU(3)$ the numerics seem compatible with $\gamma \simeq 0$ \cite{privateCom}. The rough $SU(\infty)$ estimate in this table is obtained through a simple fit of the form $a+b/N^2$ using the three data points for $N=2,3,6$. Conservatively, the green region surely contain all pure $SU(N)$ theories with $N\leq 6$.}:
\begin{table}[h]
    \centering
    \begin{tabular}{| c || c | c | c | c | }
\hline
 Gauge Group  & $\mathbb{Z}_2$  & $SU(2)$ & $SU(6)$ & $SU(\infty)$  \\
       \hline
       \hline
  $\gamma  $  & $-0.4$\,  \cite{Baffigo}
    & $-0.3$\, \cite{Caristo:2021tbk} & $+0.2$ \, \cite{dressing,Dubovsky:2014fma} & $+0.3$ \\
        \hline
    \end{tabular}
    \label{tab:my_label}
\end{table} \\
Once we impose this conservative upper bound on $\gamma$ we see that the allowed space shrinks by several orders of magnitude leading to a tiny green rock, the smallest rock in our Branon matrioska, see right panel in figure \ref{MatrioskaFig}.

\section{Discussion}
\lettrine{W}{hat} class of string theories govern the flux tube of pure three-dimensional Yang-Mills theories? What is the world-sheet physics of this two-dimensional theory at finite energies? We still don't know.

Nonetheless, we find it quite remarkable that the green region is so small that we can already predict finite energy sum rules up to a few digits for any reasonable confining gauge theory such that $SU(3)$ pure glue! While the Matrioskas in figure \ref{MatrioskaFig} were generated through primal numerics using Perimeter Institute's computer cluster but the power of the Bootstrap in constraining these sum rules can be beautifully seen analytically as well; in appendix \ref{analyticalbounds} we derive several simple single component analytic bounds including for example the sum rule bound $$0.7778 \le X \le 0.7796 \,,$$ 
for the width of the smallest branon Matrioska. 
As far as we know, the S-matrix bootstrap is the only available machinery that allows us to quantitatively address such interesting Lorentzian quantities. 

\begin{figure*}[t]
\includegraphics[width=\textwidth]{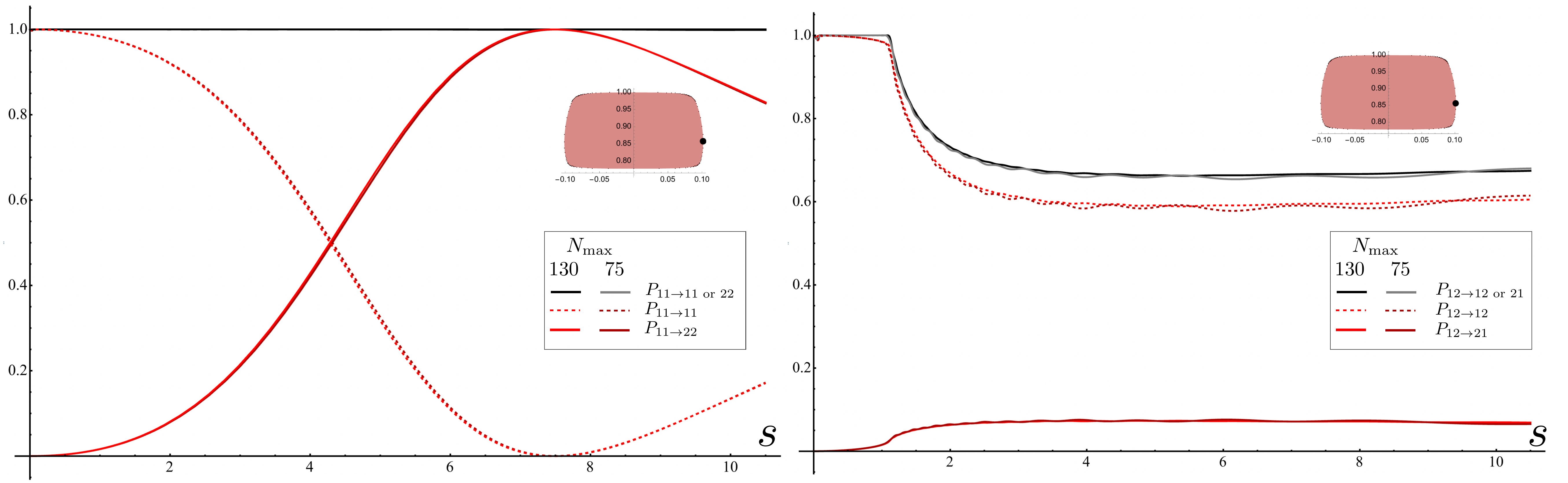}
  \caption{At very low energy, the diagonal processes $11\to 11$ and $12\to 12$ dominate since the theory is free in the IR. As we crank up the energy jet production kicks in. On the even sector depicted on the left we see that around $s=4$ the jet production $11\to 22$ even starts dominating for this rightmost point of the red Matrioska. The solid curve on that left panel at $P=1$ is simply the sum over the two possible even outcomes. On the right we have the odd sector and that same sum no longer adds up to $1$! That means that around $s=1$ the optimal S-matrices at this point of the Matrioska choose to \textit{produce} some odd state outside the $12$ branon/jet system! Would be fascinating to find out what these states are in case they have a physical meaning. This is the first instance of an S-matrix bootstrap where unitarity is perfectly saturated in a finite energy interval after which it is not. In all other examples we know of unitarity wants to converge to $1$. Note that in these plots we depicted two very different values of (very large) $n_\text{max}$ to be sure that there is no convergence issue.} \label{shiftsBoundary}
\end{figure*}

In figure \ref{shiftsBoundary} we looked for the probability for various production in the even and odd channels for the optimal S-matrices on the boundary of the two dimensional section of the red region. On the left (even) panel we observe something rather cute albeit unsurprising while on the right (odd) panel we observe something rather striking and  unexpected (based on all previous S-matrix bootstrap studies). We find that after some critical energy, unitarity is \textit{not} saturated in the odd sector. The probability of finding an odd branon plus an even jet in the final state moving as in the initial state ($P_{12\to 12}$) or reflected with respect to the initial state ($P_{12\to 21}$) do not add up to $1$! Something else is being produced. What is it? Would be fascinating - albeit probably far fetched - if they could be thought of as glue balls emitted from the flux tube when the branons reach a critical energy roughly corresponding to their mass.\footnote{There is a more conservative possibility of attaining unitarity saturation once other jet states (\ref{jet2}) are included in the bootstrap system. Although it should be checked explicitly, our expectation is that including additional jet states will not affect the results of figure \ref{shiftsBoundary}, since the numerics should just ``turn off" higher jet components at intermediate energies in the red ``X-Y" projection slab.} This would make nice contact with the upcoming explorations in  \cite{AndreaJoaoAditia}.
Even if they turn out to be more exotic objects or to have no simple physical interpretation it is still mathematically fascinating that unitarity is not saturated along the boundary.\footnote{The only other example of unitarity non-saturation that we know of is \cite{Monolith} where two very special isolated points at the boundary of the space of allowed bound-state free gapped S-matrices with $O(N)$ symmetry were identified. At these special points, the optimal S-matrices are constant and do not saturate unitarity. 
In higher dimensions it is always hard to rule out particle production since the extrapolation to infinite $n_\text{max}$ is much more involved there. There is is often hard to distinguish numerical error and convergence issues from real tiny particle production. This is not what we have here at all. Here particle production is not a small effect!, it is a big stable bump which does not seem to go away at all as we increase the number of parameters in our numerics.}  
What is going on? Could something like this also happen in higher dimensional scattering?\footnote{We know particle production is mandatory in higher dimensions. It is a theorem \cite{Aks:1965qga}, see \cite{Correia:2020xtr} for a modern review. The issue is that this theorem forbids absence of particle production but fails to put a quantitative lower bound on how much particle production one must have. Inelasticity has been recently discussed in \cite{Tourkine:2021fqh,Tourkine:2023xtu} in terms of contraction maps. Would be nice to make contact with the bootstrap in some example where particle production arises dynamically as in this case. 
} We conjecture that the answer is yes and that with a similar $Z_2$ scattering problem involving an odd and an even particle in higher dimensions we should observe similarly striking dips if we choose to probe the scattering in this system by a similar triplet kind of effective couplings; would be fascinating to check that this is the case.

But even in two dimensions there is still much to do.

We could consider infinitely many more jets, explore other directions in the infinite dimensional space of multi-particle S-matrices, and check whether the Matrioska shrinks further. We could also study flux tubes in higher dimensional confining theories. In $D=4$ -- the real world -- jets could be made out of any combination of the two branons corresponding to the two transverse directions to the flux tube.
In \cite{FluxTube}, when exploring the allowed space of leading Wilson coefficients, it was found that the boundary of the allowed space of flux tube S-matrices in $D=4$ is much more interesting than in $D=3$; 
the extremal S-matrices contain a world-sheet axion resonance whose mass and coupling match both the lattice measurements, and the estimate coming from integrability \cite{Dubovsky:2015zey}.
In \cite{Gaikwad:2023hof}, it has been conjectured that the axion contributes nonpertubatively to cancel the universal particle production\footnote{This universal contribution cancels in $D=3$ for kinematical reasons. As a consequence of this term, particle production starts earlier in the EFT expansion for $D=4$, that is  $P^\text{4D}_{2\rightarrow4} = O(s^6)$ versus $P^\text{3D}_{2\rightarrow4} = O(s^8)$, and should thus give rise to larger effects than considered in this work.} induced by the Polchinski-Strominger term, thus enhancing approximate low energy integrability \cite{Polchinski:1991ax}.
It would be interesting to check this mechanism by employing the full non-perturbative multi-particle Bootstrap.
It is thus only natural to expect that the $D=4$ Branon Matrioska will probably be even richer than the $D=3$ Matrioska studied here!

It would also be interesting to explore other interesting corners of the Matrioskas. Are there interesting theories that maximize non-diagonal scattering for instance? (Be them flux tube or not.) These could be thought of as maximally non-integrable theories. 

What about higher dimensions? Can we attack multiparticles there? We must. Otherwise, how can we hope for the non-perturbative S-matrix bootstrap to rise to the standards of perturbative quantum field theory? 
Can we find a way to tame the intricate higher dimensional singularities in a somehow controllable way? A key tool in the two-dimensional explorations we initiated here was the introduction of jets, {{which allowed us to derive bounds on the multi-particle S-matrix through exploring single variable analyticity and crossing, thus evading having to deal with the intricate analytic structure of the full multi-particle amplitude}}. In higher dimensions, can we develop a jet-effective-field theory where we expand around highly aligned particles that will have a small effective mass and might possibly be treated as effective single particles? Seems like a direction worth exploring -- see~\cite{Becher:2014oda} for a review of real-world jets.

We have at last managed to put the foot in the door of the elusive multi-particle realm, previously unexplored within the S-matrix bootstrap. Hopefully, we are able to open it wide and sneak into this unexplored world and learn big things about non-perturbative quantum fields.

\begin{acknowledgments}
We are specially grateful to Joao Penedones for initial collaboration on this project and for numerous enlightening discussions and valuable comments on the draft. We thank Simon Caron-Huot, Sergei Dubovsky, Hofie Hannesdottir, Davide Gaiotto, David Gross, Victor Gorbenko, Joan Elias Miró, Sebastian Mizera and Riccardo Rattazzi for useful discussions.
Research at the Perimeter Institute is supported in part by the Government of Canada through NSERC and by the Province of Ontario through MRI. This work was additionally supported by a grant from the Simons Foundation (Simons Collaboration on the Nonperturbative Bootstrap \#488661) and ICTP-SAIFR FAPESP grant 2016/01343-7 and FAPESP grant 2017/03303-1. AG is supported by the European Union - NextGenerationEU, under the
programme Seal of Excellence@UNIPD, project acronym
CluEs.

\end{acknowledgments}

\bibliography{refs}

\clearpage

\renewcommand\appendixname{}
\appendix

\begin{center}
\textbf{\Large Supplementary Materials}    
\end{center}

\counterwithin*{equation}{section}
\renewcommand\theequation{\thesection\arabic{equation}}
\renewcommand{\thesection}{S\arabic{section}}
\renewcommand{\theequation}{S\arabic{section}.\arabic{equation}}
\renewcommand{\thefigure}{S\arabic{figure}}
\section{Analyticity, Crossing, and All That}
\label{theoryappendix}

\subsection{A Jet as One More Particle. (Or More.)}

\begin{figure}[t!]
  \includegraphics[scale=0.3]{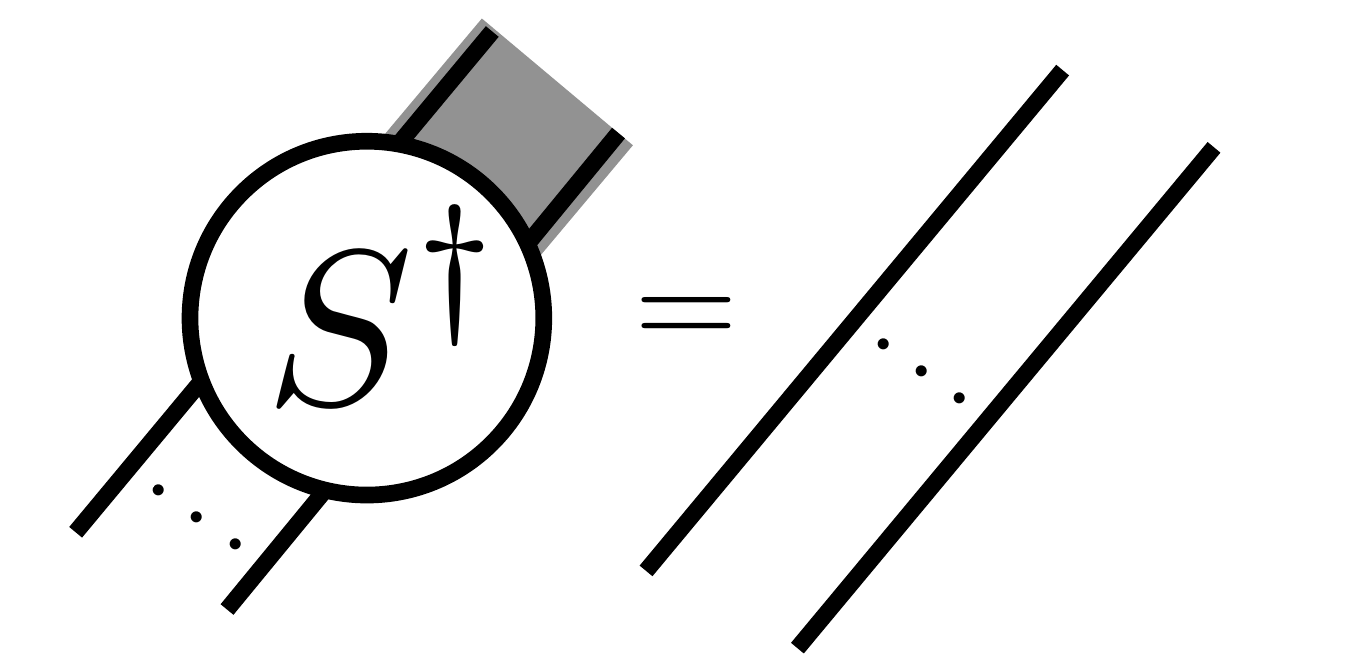}
  \caption{Particles of the same chirality do not scatter amongst each other. Indeed, we can always boost such configuration so that all particles would have very low energy and thus be effectively free.} \label{trivialLL}
\end{figure}

We defined a jet as a set of massless branons all moving collinearly.
 For jets made out of two particles (one with an energy fraction $\alpha$; the other ($1-\alpha$)) we had (\ref{jet2}), see section (\ref{higherjetsection}) for the $N$-particle case. These definitions provide good asymptotic scattering states because the constituents of these jets have trivial scattering amongst themselves  (\ref{LLRR}), see figure \ref{trivialLL}.

Kinematically, the jets are indistinguishable from single particle states. Their $2\rightarrow 2$ S-matrix elements should thus satisfy the standard analyticity, crossing and unitarity properties of $2\rightarrow2$ scattering amplitudes. For some of us this is a very natural working assumption.

For some people, however, this could be surprising. After all, the jets scattering amplitudes are really multi-particle processes, which we were taught to fear. This appendix is aimed at bringing some peace to those more timorous skeptics\footnote{Including some of the authors.}.

Perhaps the most surprising claim is that of analyticity. Aren't we supposed to have a myriad of Landau singularities in multi-particle scattering? In principle, even assuming that the amplitude is analytic could be wrong, as we expect the scattering amplitude to be the sum of boundary values of several analytic functions \cite{Bros:1972jh,Iagolnitzer:1977sw,Iagolnitzer:1994xv,Hannesdottir:2022bmo}. We will try to address all these concerns in order.

\subsubsection{Searching for Landau Singularities}
\label{landausection}

For example, let us search for possible non-trivial solutions to the Landau equations \cite{Landau:1959fi} for a generic diagram. Since we will be making an argument for all diagrams, it is enough to consider their \textit{leading} singularity, defined by having all internal lines on-shell: $q_j^2 = 0$ for internal edge $j$. Note that we are agnostic whether the respective Schwinger parameter $\alpha_j$ is non-zero, see \cite{Prlina:2018ukf, Fevola:2023kaw}. 

We will argue that either: (1) the external momenta must cluster into individually conserved subsets, or (2) the diagram corresponds to a normal threshold, which is not of concern. To do so, note that, because internal edges are on-shell massless particles in two dimensions, they must be either left or right movers. Suppose we remove all left-moving internal edges from the diagram. There are two possibilities: either the diagrams breaks apart into disconnected components or not. 

\begin{figure}[t!]
  \includegraphics[width=0.5\textwidth]{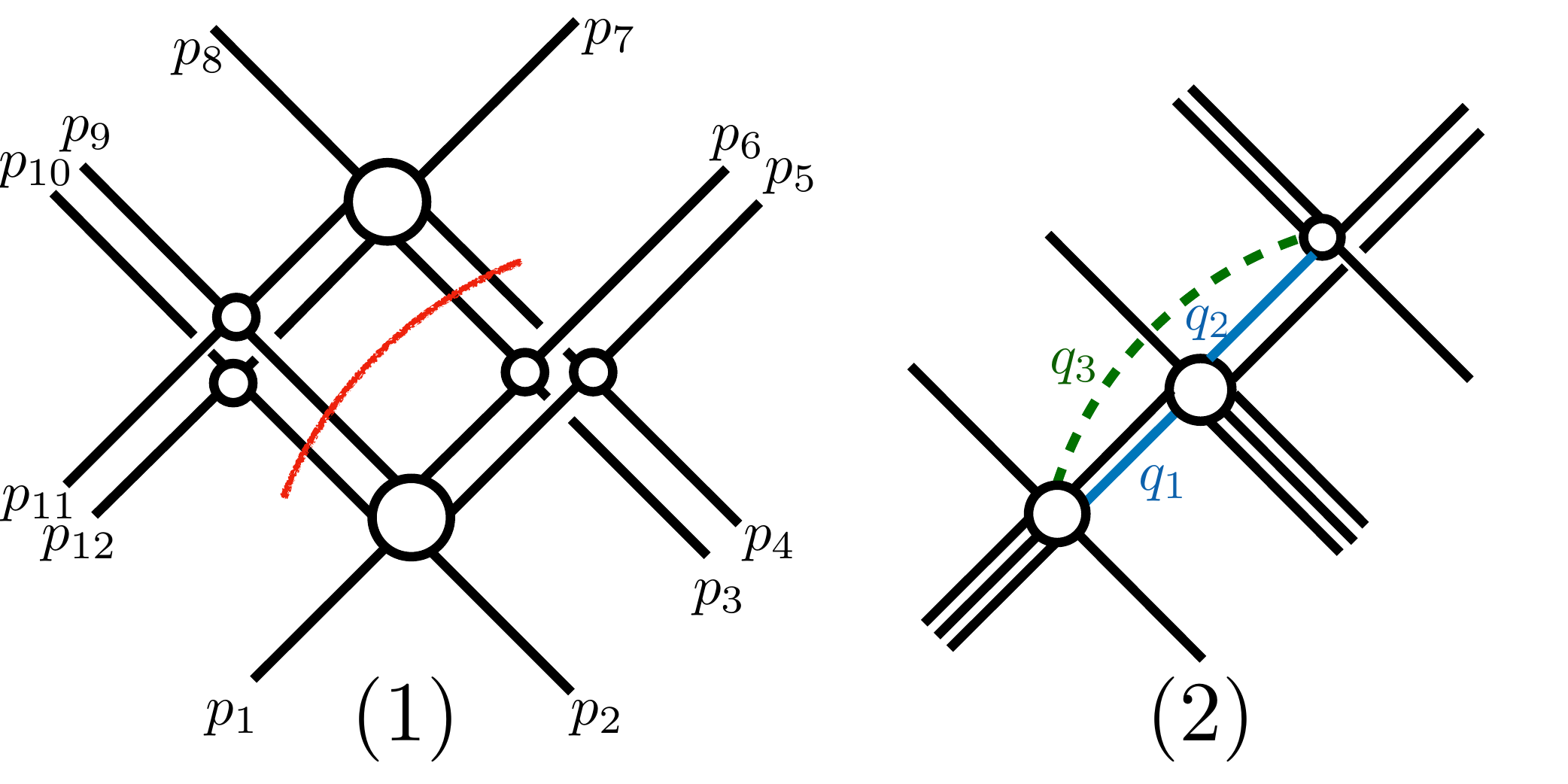}
  \caption{Leading Landau singularities come from either diagrams in which momentum cluster into subsets that scatter independently, such as the diagram in figure (1), or normal thresholds, such as the diagram in figure (2). In figure (2), $q_3$ is a left-mover, while $q_1$ and $q_2$ (and the remaining internal lines) are right movers. The loop equations imply $\alpha_3 q_3 =0$.} \label{landaufigure}
\end{figure}

In the first case, see figure \ref{landaufigure}(1) for an example, the right-moving external momenta in each disconnected component is a proper subset of the right-moving momenta of the original diagram\footnote{If one subset were empty, the respective component would have only left-moving external momenta.  Since in the theory under consideration there is no pure left-left scattering, see \ref{trivialLL}, we expect this case to have vanishing discontinuity, and hence can be disconsidered. The triangle diagram would be an example.}. Hence, by energy momentum conservation, the right-moving momenta in each component must be separately conserved, thus clustering. 

In the second case, see figure \ref{landaufigure}(2) for an example, every pair of vertices $(x_j, x_k)$ must be connected by a path of right-moving edges $\{(\alpha_i, q_i)\}$, since the diagram remains connected after removing left-movers. All vertex displacements $x_j-x_k = \Sigma_i \alpha_i q_i$ are thus collinear and the Landau graph is one dimensional\footnote{Note that, given a left-moving edge $(\alpha_j,q_j)$ in this diagram, one can complete this edge to a loop with right-moving edges $\{(\alpha_i, q_i)\}$. The Landau loop equation $\alpha_j q_j + \Sigma_i \alpha_i q_i = 0$ then implies that $\alpha_j q_j = 0$.}.
This is, by definition, a normal-threshold.

Hence we expect \textit{ no anomalous singularities at generic kinematics}. On the other hand, when the momentum clusters, we do expect anomalous threshold. It is interesting to revisit the origin of these clustering singularities. They are closely tied to causality, and manifest the familiar statement that the amplitude corresponds to the sum of boundary values of analytic functions from opposite sides. The discussion that follows is inspired by Iagolnitzer's nice paper \cite{Iagolnitzer:1977sw}, see also \cite{Chandler:1968daw, Iagolnitzer:1978my, Iagolnitzer:1994xv}.

\subsubsection{Clustering Momenta and Causality}
\label{causalitysection}

\begin{figure}[t!]
  \includegraphics[width=0.4\textwidth]{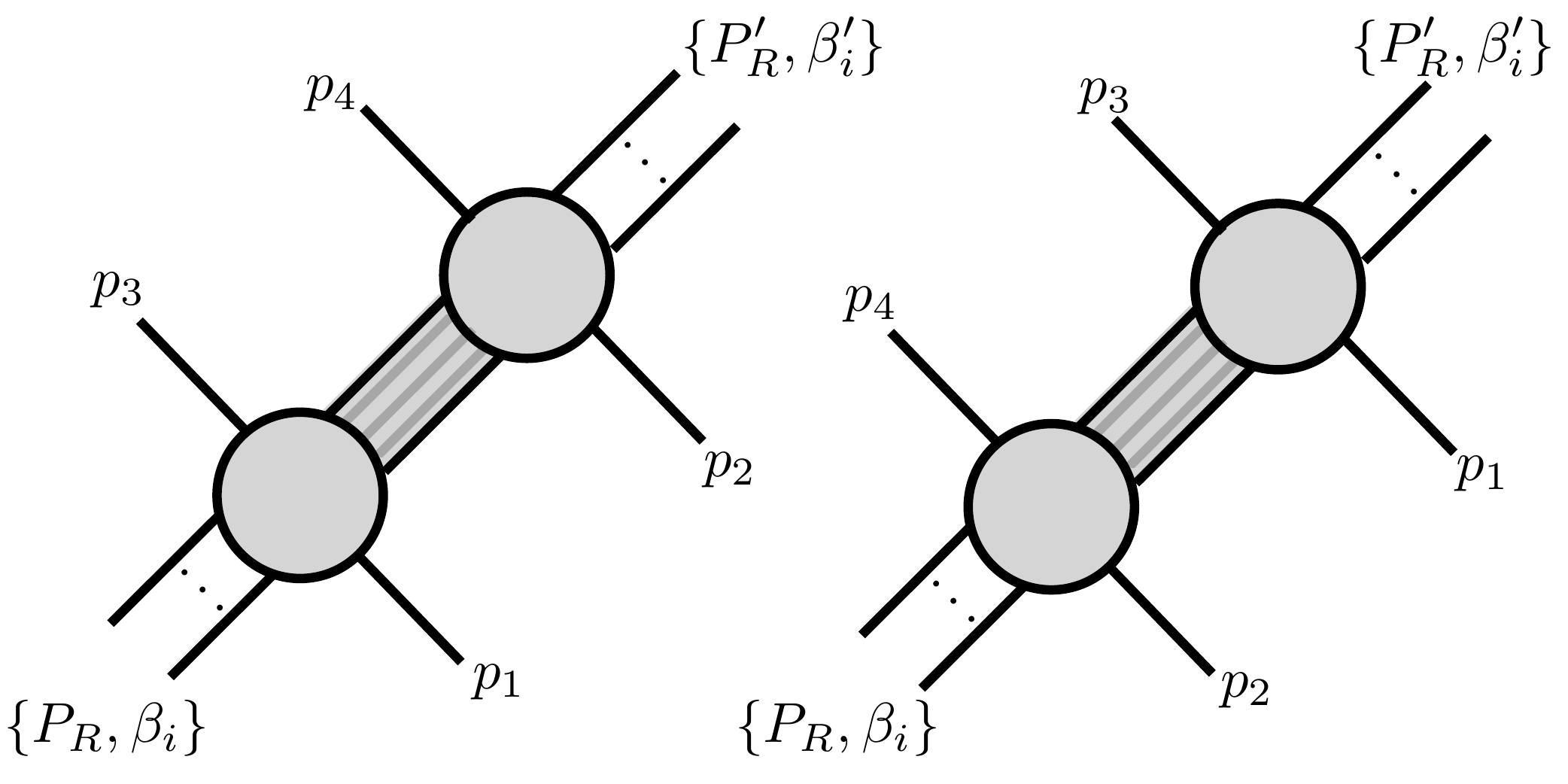}
  \caption{Diagrams in which the momentum clusters come in groups related by different causal orders of the scattering process. Correspondinly, these come with opposite causal directions in momentum space, resulting in non-analyticities that prevent continuation of individual momenta. Analyticity in the total energy is unobstructed.} \label{clusteringfig}
\end{figure}

For concreteness, consider the scattering depicted in figure \ref{clusteringfig}, in which left-movers $p_1 = \alpha P_L$ and $p_2 = (1-\alpha) P_L$ scatter against right-movers $(P_R, {\beta_i})$, producing another bundle of right-movers $(P'_R, {\beta'_i})$ and outgoing left-movers $p_3 = \alpha' P'_L$ and $p_4 =(1-\alpha') P'_L$. As argued above, we expect singularities from the diagrams in figure \ref{clusteringfig} whenever $p_1$ = $p_3$ (or $p_1 = p_4$). We see immediately that such diagrams come in pairs, related by the causal ordering in which the scattering happens. The claim is that, from causality, these diagrams have singularities with opposite causal directions, so that
\begin{align}
& T(p_L, p_R, \alpha, \alpha', {\beta_i, \beta'_i} ) \label{macrocausality}=\\ &  \left(\frac{ D_1(p_L, p_R, \alpha', {\beta_i, \beta'_i})  }{\alpha - \alpha' + i \epsilon} +  \frac{ D_2(p_L, p_R, \alpha', {\beta_i, \beta'_i})  }{\alpha' - \alpha + i \epsilon} + \texttt{reg} \right) \nonumber,
\end{align}
where $D_1$ comes from the first set of diagrams, $D_2$ from the second set of diagrams and $\texttt{reg}$ is a background analytic at $\alpha = \alpha'$. $D_1$ and $D_2$ may contain additional singularities, e.g. when the right movers also cluster. Note that this implies that the amplitude is the sum of boundary values of analytic functions from opposite directions from the point of view of the energy fractions, but analyticity on the total energy is not obstructed.

To see how (\ref{macrocausality}) can be inferred from causality, consider integrating the external momenta $p_1$ against the following wave function:
\beq
\int dp_1  e^{-|\tau| (p_1- p_3)^2 - i \tau p_1}.
\eeq
At large $\tau$ we see, from the stationary phase, that, in the asymptotic past, this corresponds to the particle moving with (spatial) momenta $-p_3$ along the classical trajectory $x = - p_3  t - \tau $. In other words, we translate the trajectory of particle $p_1$ far to the left. Causality then suggests that the diagrams $D_1$ should dominate the scattering process. Indeed, we see that, deforming the integral contour slightly into the lower half plane around $p_1 = p_3$ leads to an exponential suppressed integral, up to the residue contribution $D_1$. If, instead, we set $\tau\rightarrow-\tau$, so that we translate particle $p_1$ to the far right, we expect diagrams $D_2$ to contribute, and, indeed, we must now deform into the upper half plane thus picking the pole in the second term of (\ref{macrocausality}).

\begin{figure}[t!]
  \includegraphics[width=0.4\textwidth]{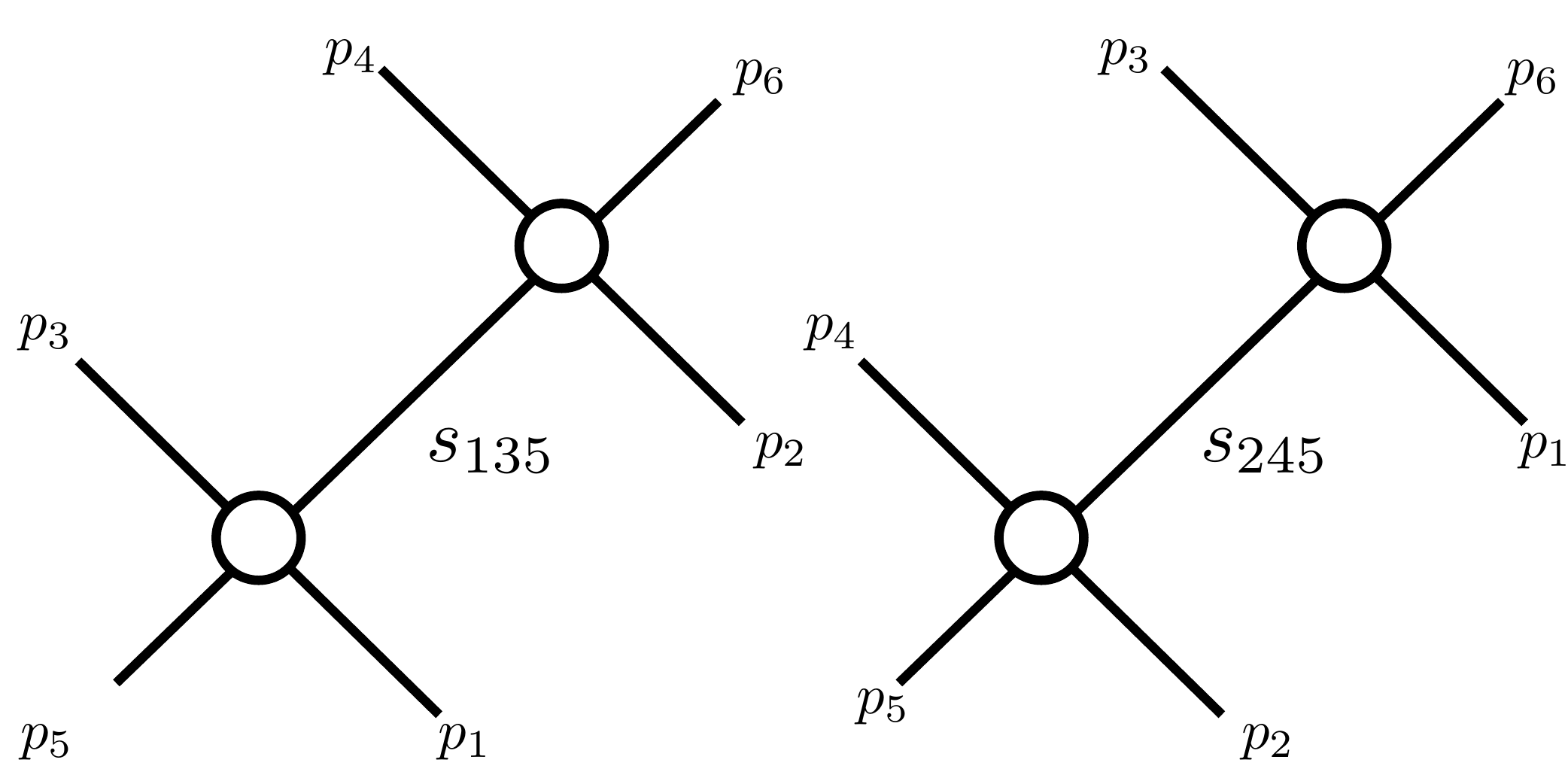}
  \caption{A tree level example manifesting the structure of equation (\ref{macrocausality}). It results in the contribution (\ref{treeequation}).} \label{treefig}
\end{figure}

Note that this discussion applies already to the tree level scattering. Consider the scalar exchange  of figure \ref{treefig}. In the physical region\footnote{In the physical region we have $s = P_L P_R > 0$ but $s_{135}$ and $s_{124}$ have no definite sign.}  the diagrams contribute
\beq
 \frac{i}{s_{135} + i \epsilon } +   \frac{i}{s_{245} + i \epsilon } = \frac{2 \pi}{s} \delta(\alpha - \alpha') \label{treeequation}
\eeq
which manifest the structure (\ref{macrocausality}) with $\texttt{reg} =0$. 
Again, note that, upon integrating the multi-particle S-matrix against jet wavefunctions, we have perfect analyticity in $s$ as well as standard crossing symmetry 
$M_{{12\rightarrow12}}(-s^*) = M_{{12\rightarrow12}}^*(s)$ from these contributions.

Let us also mention that equation (\ref{macrocausality}) is central to the possibility of integrable models in two dimensions, for which, e.g, the $\text{left-right}\rightarrow \text{left-left-left-right}$ amplitude vanishes but the $\text{left-left-right}\rightarrow \text{left-left-right}$ amplitude is non-zero and, in fact, given by a factorized product of $2\rightarrow 2$ S-matrices \cite{Iagolnitzer:1977sw}. What happens in that case is that \beq
D_1 = D_2 =  (i M_{2\rightarrow2}(s_{12})) (i M_{2\rightarrow2}(s_{13}))/s\label{yangbaxter}\eeq and $\texttt{reg} = 0$ so that
\beq
S_{3\rightarrow3} = S_{2\rightarrow2}(s_{12}) S_{2\rightarrow2}(s_{13}).
\eeq
The $\delta$-function $\delta(\alpha-\alpha')$ prevents crossing from $3\rightarrow3$ (which is non-zero) to $2\rightarrow4$ (which is zero) kinematics, see also the review \cite{Dorey:1996gd} for a discussion. An example of a massless integrable S-matrix is the scattering theory describing the RG-flow from the tricritical to critical Ising model \cite{Zamolodchikov:1991vx}. It serves as a non-perturbative example in which to test the various claims in this section, albeit in a trivial way. In integrable theories for which left-left and/or right-right scattering are non-trivial\footnote{For examples see \cite{Zamolodchikov:1992zr,Fendley:1993xa}.}, or in massive integrable theories, and in theories with flavor indices, equation (\ref{yangbaxter}) gets modified by and extra scattering factor, and the condition $D_1 = D_2$ becomes the standard Yang-Baxter equation.

\subsubsection{Crossing Symmetry}
\label{crossingsection}

\begin{figure}[t]
  \includegraphics[scale=0.14]{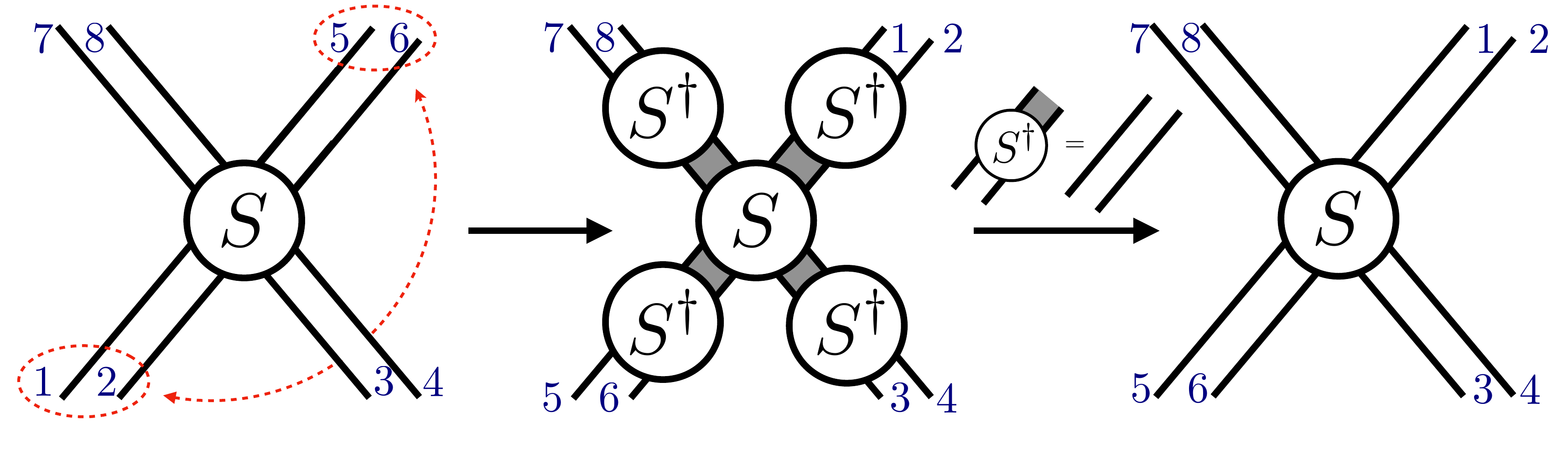}
  \vspace{-0.2cm}
  \caption{Crossing of two particles from past to future would generically give rise to an amplitude dressed by several external S-matrices. Here all these S-matrices are trivial since they involve particles of the same chirality.} \label{crossingjets}
\end{figure}

Crossing is also expected to be non-trivial in multi-particle amplitudes. Consider for example the elastic scattering between two 2-branons jets. Each jet in the past or future is made of two branons so we can think of this elastic process as a scattering event involving a total of eight fundamental particles. Swapping a jet in the future with a jet in the past is thus equivalent to swapping 4 of these 8 particles as indicated on the left of figure \ref{crossingjets}. A beautiful recent paper \cite{Caron-Huot:2023ikn} studied crossing in multi-particle processes. As explained there, at least under favourable circumstances\footnote{As argued above, the branon S-matrix should also be a favourable case, at least for the crossing move discussed here, in which we cross the total energy, and not sub-energies.}, what we should expect is that such a crossing transformation in an eight-particle scattering amplitude leads to a slightly more involved object where an eight-particle amplitude is conjugated by a bunch of external S-matrices as indicated by the diagram in the middle of figure \ref{crossingjets}.\footnote{As explained in \cite{Caron-Huot:2023ikn} this general expectation reduces to the usual trivial crossing transformation for usual $2\to 2$ scattering since such external S-matrix dressings, in that case, would involve a single leg and stability of the external particles simply translates into $S|\texttt{single particle}\rangle=|\texttt{single particle}\rangle$ so that those extra factors are gone again.} But here, since the scattering inside each jet is trivial, figure \ref{trivialLL}, we can simply wash away those extra external scattering factors to obtain the crossed amplitude with the jets swapped as depicted on the right of figure \ref{crossingjets}.

\section{Normalization, Unitarity and Probabilities}
\label{unitaritysection}

We can treat the massless jet in (\ref{jet2}) as a fundamental even particle with respect to crossing and analyticity of their S-matrix elements. Note that the various factors in its definition also ensure a proper single-particle normalization. 

We define the \emph{jet-projector} for a two-particle state as
\beq
\mathbb{P}_n^\alpha[f(\alpha)]=\sqrt{\frac{2n+1}{8\pi}}\int\limits_0^1 d\alpha \frac{P_n(2\alpha-1)}{\sqrt{\alpha(1-\alpha)}}f(\alpha).\label{2particle_jet_projector}
\eeq

We defined the two co-linear particles jet with momenta $p_1=\alpha p$, $p_2=(1-\alpha)p$ as 
\beq \label{choicebasis}\ket{n,p}=\mathbb{P}_n^\alpha[\ket{\alpha,1-\alpha,p}].\eeq
From there it follows that, for even $n$,
\beqa
\braket{n,p|n^\prime,p^\prime}&=&16 \pi^2 p \delta(p-p^\prime) (\mathbb{P}_n^\alpha \circ \mathbb{P}_{n^\prime}^{\alpha^\prime}) \nn\\
&&[\alpha(1-\alpha)(\delta(\alpha-\alpha^\prime)+\delta(\alpha-1+\alpha^\prime))]\nn \\
&=& 4\pi p \delta(p-p^\prime)\delta_{n n^\prime}.
\label{norm_jets}
\eeqa
which is the standard one-particle normalization.

 We also have the completeness relation
\beq
\frac{1}{2}\int \frac{dp_1}{4 \pi p_1} \frac{dp_2}{4 \pi p_2} |p_1 p_2 \rangle \langle p_1 p_2 |  = \sum_{k=0}^\infty \int \frac{d P}{4 \pi P} |2k, P\rangle \langle 2k, P | \label{complete}
\eeq

Define here the single branon state as $|P\rangle \equiv |-2 , P\rangle$ so that we have uniform notation for both single and two particle jet states. The projector into one left- and one right- as well as two left- and two right- movers subspace is thus given by
\beq
\sum_{\substack{k=-1 \\ k'=-1}}^\infty \int \frac{dP_L}{4 \pi P_L} \frac{dP_R}{4 \pi P_R}  |2k, P_L\rangle\otimes|2k', P_R\rangle \langle 2k, P_L| \otimes \langle 2k', P_R|.
\label{projector} \eeq

Projecting intermediate states in the unitarity equation $$\langle n, P_L| \otimes \langle n', P_R|(\mathbf{1} = \mathbf{S} \mathbf{S}^\dagger)|m, P'_L\rangle\otimes|m, P'_R\rangle $$ with (\ref{projector}) results in
\beq
\mathds{1} \succcurlyeq \mathbb{S}.\mathbb{S}^\dagger \label{unitaritysemi}
\eeq
where 
\begin{align*} \mathds{1}_{(n n')(m m')} &= \delta_{(n n')(m m')}\\ \mathbb{S}_{(n n')(m m')} &= S_{n n' \rightarrow m m'}(s)
\end{align*}
and
\begin{align*}
&\langle n, P_L| \otimes\langle n', P_R| \mathbf{S}|m, P'_L\rangle\otimes|m, P'_R\rangle  =\\ & 16 \pi^2 P_L P_R \delta(P_L - P_L')\delta(P_R - P_R') S_{n n'\rightarrow m m'}(s)
\end{align*}

In particular $n = n' = m = m' = -2$ results in (with main text notation)
\beq
1 \geq |S_{11 \to 11}|^2 + \sum_{m',m}|S_{11\to m m'}(s)|^2,
\eeq
which allow us to identify
\beq
P_{2\to4} = \sum_{m',m}|S_{11\to m m'}(s)|^2
\label{production_with_jets}
\eeq
as the probability for particle production in the sub-sector considered\footnote{That is, $LR\rightarrow LLRR$, which is not all four-particle production since e.g. $LR \rightarrow LLLR$ is non-zero.}.

In S-matrix bootstrap computations, one can consider the submatrix of (\ref{unitaritysemi}) with $n,n',m,m'\leq N_{max}$ in order to obtain a finite set of unitarity constrains to be imposed numerically. One should also project into intermediate states with $2k, 2k' \leq N_{\max}.$ In this paper, we used $N_{max} =0$.

\section{Higher jet states}
\label{higherjetsection}
Let us now describe a basis for n-particle jet states. This can be used to extend the multi-particle bootstrap initiated here to the full $n\rightarrow m$ scattering amplitude. Note that the analyticity and crossing discussion above was general. 

A $\texttt{basis}$ $|\texttt{b}_N(\vec{n})\rangle$ for $N$ particle jets parametrized by the multi-index $\vec{n}$ can be constructed starting from the ansatz
\begin{align}
&|\texttt{b}_N(\vec{n}) \rangle \equiv 2^N \int_0^1  \left(\prod_{i=1}^N\frac{d\alpha_i}{\alpha_i^{\beta}}\right)\frac{f_{\vec{n}}(\alpha_i) \delta\left(1- \Sigma_i \alpha_i\right)}{\sqrt{(4 \pi)^{N-1} N!}} | \alpha_i P\rangle \nonumber
\end{align}
with $f_{\vec{n}}$ a basis of polynomials and $\beta<1$ so that the state is square integrable with the on-shell measure 
$d\mu(p_i) =\left(\prod_i dp_i/(4 \pi p_i) \right)$.
Since the particles are identical, it is enough to consider symmetric polynomials.

In the case $N=2$, equation (\ref{choicebasis}), we chose $\beta = 1/2$ in which case an orthonormal basis is provided by the Legendre polynomials. This choice of $\beta$ is not unique and was made for convenience. 

Let us consider the inner product of two elements. We get
\begin{equation}
\hspace{-0.5cm}\frac{\langle \texttt{b}_N(\vec{n}) | \texttt{b}_N(\vec{m}) \rangle}{ 4 \pi P \delta(P - P') } =\int\limits_0^1  \left(\prod_{i=1}^N\frac{2 d\alpha_i}{\alpha_i^{2\beta-1}}\right) \frac{\delta\left(
1-\Sigma_i \alpha_i\right)}{\left(f_{\vec{n}}(\alpha_i)f_{\vec{m}}(\alpha_i)\right)^{-1}}
, \nonumber
\end{equation} 
where we used permutation symmetry of the polynomials to relate all Wick contractions. Changing variables as
$x_i^2 = \alpha_i$, we get
\begin{align}
\texttt{lhs} =2^N \int_0^1  \left(\prod_{i=1}^N\frac{dx_i}{x_i^{4\beta-3}}\right)f_{\vec{n}}(x_i^2)f_{\vec{m}}(x_i^2) \delta\left(1- \Sigma_i x_i^2\right), \nonumber
\end{align} 
It is now convenient to take $\beta = 3/4$ so that, using $x_i \to -x_i$ symmetry to extend the range of integration to $(-1,1)$, we finally get
\begin{align}
\frac{\langle \texttt{b}_N(\vec{n}) | \texttt{b}_N(\vec{m}) \rangle}{ 4 \pi P \delta(P - P') } =\int_{S_{N-1}} dS_{N-1} f_{\vec{n}}(x_i^2)f_{\vec{m}}(x_i^2),
\end{align} 
where we now integrate over the sphere with the standard surface measure. An orthonormal basis (with respect to the surface measure) for square integrable functions on the sphere is provided by the harmonic polynomials (spherical harmonics). In our case, it is enough to consider symmetric harmonic polynomials of even degree in each variable $x_i$. These can be easily constructed recursively (in the overall degree of homogeneity) from the spherical harmonics and linear algebra. Once that is achieved, we have
\beq
\langle \texttt{b}_N(\vec{n}) | \texttt{b}_N(\vec{m}) \rangle =  4 \pi P \delta(P - P')  \delta_{\vec{n}\vec{m}}.
\eeq

It follows that we have the completeness relation for states of  $N$ collinear particles as
\beq
\int \left(\prod_{i=1}^N \frac{dP_i}{4 \pi P_i}\right) |P_i\rangle \langle P_i| = \sum_{\vec{n}} \int \frac{dP}{4 \pi P} | \texttt{b}_N(\vec{n}) \rangle \langle \texttt{b}_N(\vec{n}) | 
\eeq

Unitarity for the full S-matrix then takes the form (\ref{unitaritysemi}) with $n, n', m, m'$ being promoted to multi-indices whose order is the number of particles in the jet.

\section{EFT and the Jet S-matrix} 
\label{apFD}

In this appendix we revert to the more conventional normalization
\beq
\gamma_\text{this appendix} = \frac{\gamma_\text{main text}}{768}
\eeq
for $\gamma$ not to clutter the formulae with ugly $768$ factors.

Following \cite{Conkey:2016qju,Dubovsky:2017cnj,dressing}, it is useful to express the flux-tube S-matrix as a universal ``gravitational dressing'' times a theory dependent undressed function
\beq
\mathbf{S}= e^{i \ell_s^2 P_L P_R} \mathbf{S_u},
\label{gravity_dressing}
\eeq
where $P_L$ and $P_R$ are the total left and right momenta flowing into the process, and $S_u$ denotes the undressed non-universal contribution. 

For the diagonal scattering amplitudes, i.e.  those with equal numbers of left-moving and right-moving particles in the incoming and outgoing states, to first non-trivial order, $S_u$ has no connected component, since one can use at most one vertex of the generating function (\ref{lagrangian}) at this order, and the six-point vertex does not contribute to diagonal processes. It follows that, up to order $s^5$ all diagonal S-matrices have factorized scattering:
\beq
\mathbf{S}^\text{diagonal}_{3\rightarrow 3} = \mathds{1}_{3\to3} \times S_{11 \rightarrow 11}(s_{12})S_{11 \rightarrow 11}(s_{13}),\label{3to3easyparam}
\eeq
and
\begin{align}
\mathbf{S}^\text{diagonal}_{4\rightarrow 4} = \mathds{1}_{4\to4} &\times S_{11 \rightarrow 11}(s_{13})S_{11 \rightarrow 11}(s_{14})\\&\times S_{11 \rightarrow 11}(s_{23})S_{11 \rightarrow 11}(s_{24}), \nonumber
\end{align}
where
 \begin{align}
\nonumber\mathds{1}_{3\to3} =& (4\pi)^3 P_L P_R \alpha (1-\alpha) \delta({P}_L - P'_L) \delta(P_R - P'_R) \\&\times \left(\delta(\alpha - \alpha') + \delta(1- \alpha - \alpha')\right)\label{4to4easyparam}
\end{align}
and
 \begin{align}
&\nonumber\mathds{1}_{4\to4} =(4\pi)^4 P_L P_R \alpha (1-\alpha) \beta (1-\beta) \delta({P}_L - P'_L) \\& \times \delta(P_R - P'_R) \nonumber \left(\delta(\alpha - \alpha') + \delta(1- \alpha - \alpha')\right)\\&\times \left(\delta(\beta - \beta') + \delta(1- \beta - \beta')\right)
\end{align}
where $P_{L/R}$ is the total left/right moving momenta. Above, left-moving jets have incoming/outgoing energy fractions $\alpha/\alpha'$, while right-moving jets have fractions $\beta/\beta'$.

Notice that \eqref{3to3easyparam}, and \eqref{4to4easyparam} are a non-trivial consequence of the effective flux tube theory. A full direct check in perturbation theory requires computing the four-point amplitude up to two loops, and the six-point amplitude up to one-loop.

 The only non-diagonal process at this order is the $S_{LR \rightarrow LLRR}$ process (or the crossing related process $S_{LRR\rightarrow LLR}$), which can be computed using the EFT Lagrangian \cite{dressing}, resulting in 
\beq
\mathbf{S}_{2 \rightarrow 4}(s,\alpha',\beta') = \mathds{1}_{2\to2}\left(-\frac{3}{2} i\gamma \ell_s^8 \tau(\alpha') \tau(\beta')\right) s^4
\label{2to4pert}
\eeq
where $\tau(x) = x(1-x)(1-x+x^2)$ and $ \mathds{1}_{2\to2} = (4 \pi)^2 P_L P_R \delta(P_L - P'_L) \delta(P_R - P'_R) $. 

Applying the operator \eqref{2particle_jet_projector} we can project the processes \eqref{3to3easyparam}, \eqref{4to4easyparam}, and \eqref{2to4pert} into the jet basis.
The result is, up to $O(s^5)$ corrections,
\begin{align*}
S_{n1\rightarrow m1} = &\exp(i s/4)(\delta_{nm} (1 + i \gamma s^3  f_n)  + \delta_{m n\pm2}  i \gamma s^3 g^\pm_n ), \\
S_{np\rightarrow mr} = &\exp(i s/4)(\delta_{nm}  + i \gamma  s^3 \left(\delta_{nm}   f_n  + \delta_{m n\pm2} g^\pm_n \right)\\&\times\left(\delta_{pr}  f_p  + \delta_{r p\pm2} g^\pm_p \right)), \nonumber\\
S_{11 \rightarrow nm} = & -\frac{3}{2} i s^4 \gamma c_n c_m.
\end{align*}
where $g^-_x=g^+_{x-2}$ and
\begin{align}
f_n &= \frac{5n^2+5n-3}{2(2n+3)(2n-1)}\\
g^+_n &= \frac{3(n+1)(n+2)}{4(2n+3)\sqrt{(2n+1)(2n+5)}}\\
c_n &= \sqrt{\frac{2n+1}{8\pi}}\frac{\left(4 n^2+4 n-39\right) \Gamma \left(\frac{n}{2}-\frac{3}{2}\right) \Gamma \left(\frac{n}{2}+\frac{1}{2}\right)}{256 \Gamma \left(\frac{n}{2}+1\right) \Gamma \left(\frac{n}{2}+3\right)}.\label{cn_coefficients}
\end{align}

\begin{figure}
    \centering
\includegraphics[width=0.45 \textwidth]{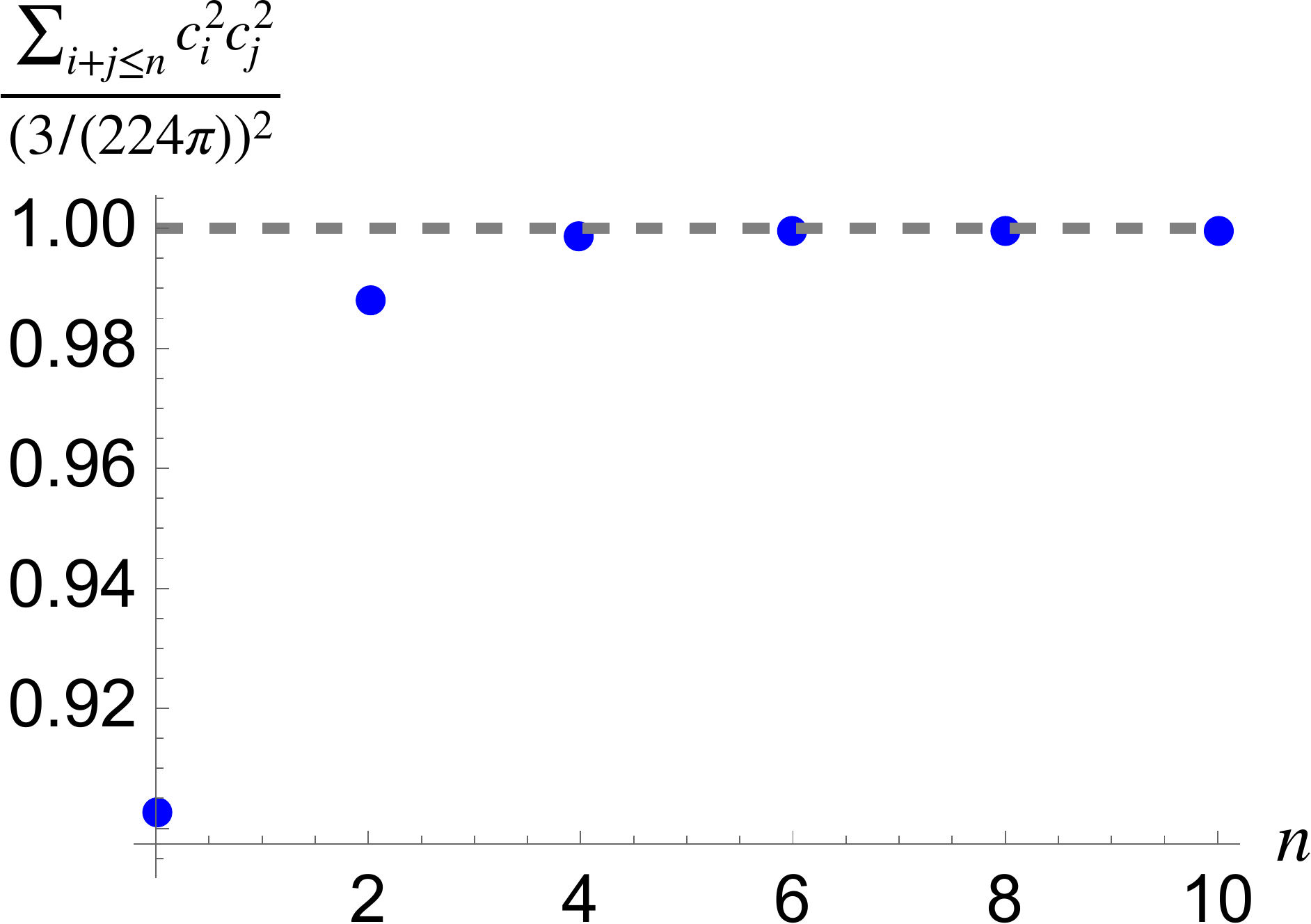}
    \caption{$P_{2\to 4}$ production probability coming from the sum of jets with $j+h\leq n$ as function of $n$ normalized by the total probability. We notice that over the 90\% of the value comes from the leading jet.}
    \label{fig:production_decomposition}
\end{figure}

Using \eqref{cn_coefficients} we can estimate how much each jet contributes to the low energy particle production. In figure \ref{fig:production_decomposition} we plot the truncated sum in \eqref{production_with_jets} normalized by the total production. We observe that the $j=0$ jet accounts for over the 90\% of the total low energy $2\to 4$ production.

\section{Good Problems for Good Solutions} \label{goodP}

Let us recall on a few trivial examples that to get a good bootstrap bound we need a good question. In all the problems below we consider  $S(s)$ to be a meromorphic function in the upper half plane which is purely real in the imaginary $s$ axis and bounded by $|S| \le 1$ in the real axis\\

  \lettrine{P}{roblem 1}. What is the max/min value of the function $S(s)$ at~$s=i$? The answer is clearly $\pm 1$ by the maximum modulus principle. The function that saturates the upper bound is the constant function $S(s)=1$ and the function that saturates the lower bound is $S(s)=-1$. \\

  \lettrine{P}{roblem 2}. How does an extra constraint $S(0)=1$ affect the bounds? The new condition has \textit{no} effect on the bound. Indeed, for the upper bound  $S(s)=+1$ is still allowed and for the lower bound we can get arbitrarily close to $S(i)=-1$ by considering the function 
\beq
S(s)=\frac{i\epsilon-s}{i \epsilon+s}  \label{dist1}
\eeq
with a very small $\epsilon$. This function is $+1$ for $s=0$ but for any $s$ of order $1$ it is very close to $-1$. Imposing $S(0)=1$ changed nothing as far as the bound goes because the solution changes very rapidly to $-1$ away from the origin completely washing out the extra constraint. This problem also teaches us another valuable lesson: the optimal solution is a function analytic everywhere in the upper half-plane, except the real $s$ axis. On the real $s$ axis the family of functions \eqref{dist1} converges for $\epsilon \to 0$ point-wise to a discontinuous function. We should keep in my mind this is the space of functions we explore when solving an S-matrix bootstrap problem. \\

 \lettrine{P}{roblem 3}. Impose now the stronger condition $S(s)=1 + i  \alpha s +O(s^2)$ where $\alpha>0$ is fixed.\footnote{
If we relax polynomial boundedness we can have theories with~$\alpha<0$. However, this would lead to superluminal signal propagation~\cite{Cooper:2013ffa}.} Since we are now imposing a fixed rate of change at the origin we can no longer easily wash out this condition and we do get an improved lower bound 
\beq
\frac{1- \alpha/2}{1+\alpha/2}  \le S(i) \le 1  \la{alphaBound}
\eeq
where the lower bound is saturated by the single CDD zero $S_{\text{CDD--}\alpha}(s)=(1+i \alpha s/2)/(1-i \alpha s/2)$ while the upper bound is untouched as we can cook up a dressing by weakly coupled resonances which satisfies the EFT behavior at threshold while relaxing back to $1$ quickly for any $s$ away from threshold: 
\beq
S(s)= \frac{  s^2-2 i   s \epsilon -4 \epsilon/\alpha }{   s^2+2 i   s \epsilon -4 \epsilon/\alpha }  \label{dist2}
\eeq
where $\epsilon$ is very small. 

In sum, a simple pointwise constraint at the boundary can lead to a big effect in the bulk as in the lower bound in (\ref{alphaBound}) or be completely washed out as in the upper bound in (\ref{alphaBound}) or the lower bound in problem 2. 
In both cases, the solutions (\ref{dist1}) and (\ref{dist2}) are best understood as distributions with a very fast changing behavior as $\epsilon \to 0 $ close to threshold  and a smooth behavior elsewhere. (We encourage the reader to plot these simple functions.) One way to forbid such singular distributions would be to impose the constraints not only point wise but in a small but finite region around the origin by imposing, say, that $|S-1-i \alpha s| < \delta$ for some $|s|<\Lambda$ for instance. In that case both upper and lower bounds will be affected but the game becomes more phenomenological as it will depend on the value of the parameters $\delta $ and~$\Lambda$ -- see \cite{EliasMiro:2022xaa,Acanfora:2023axz} for examples of such constraints in higher dimensional S-matrix bootstrap. 

In the main text we also asked if there is an improvement of the red surface where we impose not only Nambu-Goto universality but also the next few terms parametrized by $\gamma$. As mentioned there, imposing these extra point-like constraints  do nothing: the resulting problem  still seems to converge to the same red surface eventually. This should now be seen as one more example similar to the ones just discussed.

\section{Inelasticity from sum rules and a toy model} \label{toy_production}

As discussed in footnote \ref{gammabound}, the bound in equation (\ref{gamma}) is not sensitive to the presence of particle production induced by the EFT.
Here we explain why this
statement is not surprising and provide a toy model for
inelasticity that affects the allowed values of $\gamma$.

Consider the leading order particle production of the $2\to 4$ type. At leading order at low energy, using (\ref{production_with_jets}), we would conclude that 
\beqa
&&P_{2\to 4}(s)=\sum_{j,h}|S_{jh}(s)|^2=\ell_s^{16} s^8 \gamma^2 \underbrace{ \frac{1}{262144} 
\sum_{jh}c_{2j}^2 c_{2h}^2 }_{\simeq 7 \times 10^{-11}} \nn 
\eeqa
which is obviously a minuscule effect if all parameters here are of order $1$! We can estimate the effect of this little particle production on $\gamma$ by recalling the sum rule 
\beq
\frac{\gamma-2}{768}=\frac{1}{\pi}\int\limits_0^\infty \frac{\Im T_{11\to 11}(s)-\tfrac{s^3}{16}}{s^5}.
\label{sum_rule}
\eeq
so that $\delta \gamma \simeq 768/\pi \int P_{2\to 4}/s^4$ -- cut-off at some order $1$ high energy scale (could be when production reaches one or any other arbitrary scale) -- would be very tiny indeed. 

So how can different flux tubes -- all with a very similar low energy behavior -- have such different $\gamma$'s as recalled at the end of section \ref{MatrioskaS}? It must be that finite energy regions -- away from the perturbative IR regime -- are strongly contributing to (\ref{sum_rule}). In other words, non-perturbative effects in the UV completion of these theories are crucial to explain such disparate behaviors. 

Let us consider a toy model where such effects at finite energy would indeed impact $\gamma$ strongly. We imagine a situation where $S_{11\to 11}(s)=S(s)\simeq e^{i s/4+ i \gamma s^3/768}$ at low energy as predicted by EFT but $|S(s)|\le \epsilon $ for $s>s_0$.\footnote{This is possible as long as $s_0\ge -8 \log(\epsilon)/\pi$ which we assume henceforth. There would be no (polynomially) bounded, analytic,
unitarity $S$-matrix
were this condition violated. This constraint will be clear below when we derive the analytic bound (\ref{gammaAnalytic}).}\footnote{Figure \ref{shiftsBoundary} in the main text resembles such situation with $s_0 \sim 2$ and $\epsilon^2 \sim 0.7$.} 
That is we imagine a sort of black hole like model where after some critical threshold energy particle production kicks in and inelasticity dominates completely. Then we ask what is the allowed minimum value of $\gamma$ which now depends on the toy model parameters. We find a beautiful simple result\footnote{To derive this result note that $g(s)=e^{\frac{i \log(\epsilon)}{\pi} \left(\log (s_0-s)-\log
   \left(s+s_0\right)\right)}$ is analytic in the upper half plane and obeys $|g(s)|=1$ for real $s<s_0$ and $|g(s)|=\epsilon$ for real $s>s_0$. It is also real analytic, crossing symmetric, non-vanishing, and equal to one at $s=0$. It follows that the function $f(s)\equiv S(s)/g(s)$ is bounded by $1$ in the upper half plane. We also 
  know its low energy expansion up to cubic order since we know the expansion of both $S$ and $g$. Hence, following \cite{FluxTube} we can now use Schwarz-Pick analysis  for this function to bound its third derivative in terms of lower derivatives. This leads to (\ref{gammaAnalytic}). In the process, we find $f(s)=1+is/4(1+8 \log(\epsilon)/\pi s_0)+\dots$ from where we see that $\epsilon$ and $s_0$ need to obey the relation quoted two footnotes ago --  or its effective string length would be negative leading to unitarity violations.} 
\beqa
&&\!\!\!\!\!\!\!\!\!\! \gamma(s_0,\epsilon) \ge -1 \label{gammaAnalytic}\\
&&\!\!\!\!\! +\frac{-8 \log \epsilon  \left(3 \pi ^2 {s_0}^2+24 \pi  s_0 \log \epsilon
+64 \log ^2 \epsilon +64 \pi ^2\right)}{\pi ^3 {s_0}^3} \nn
\eeqa
The second line is positive so that we get a stronger bound compared to (\ref{gamma}). Note that the second line vanishes when $s_0 \to \infty$ or $\epsilon \to 1$ as expected. 

\begin{figure}[t]
    \centering
    \includegraphics[width=0.45 \textwidth]{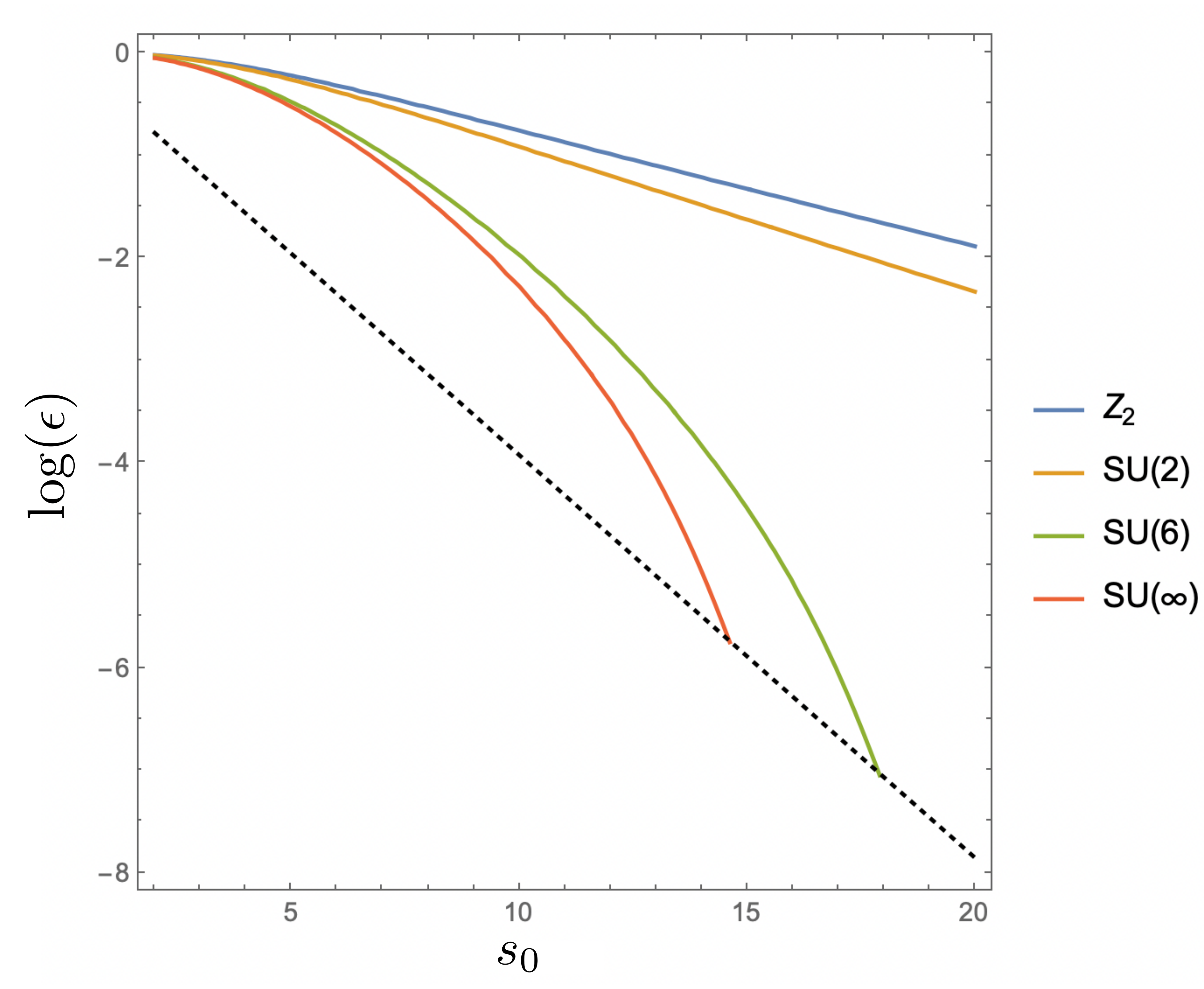}
    \caption{
    For these four solid curves we would get $\gamma$ greater or equal than $-0.4, -0.3, 0.2$ and $0.3$ which are the values mentioned in the table at the end of section \ref{MatrioskaS}. We see that to \textit{put relevant gauge theories at the boundary} we would need quite small values of $\epsilon$ and $s_0$ for large $N$ gauge theories but very mild ones for lower $N$. Perhaps for those the toy model discussed here is a good toy model.}
    \label{plotGamma}
\end{figure}

In figure \ref{plotGamma} we plot the loci $(s_0,\epsilon)$ such that the right-hand side of (\ref{gammaAnalytic}) would coincide with the various lattice results reported above. We see that indeed, for reasonable $s_0$ and not too extreme values of $\epsilon$ we could already have a big impact on $\gamma$. This is particularly true for $\mathbb{Z}_2$ or $SU(2)$ gauge theories where we expect emission of glueballs above the threshold $s_0=m_G^2\simeq 16$. For large $N$ gauge theories, however, we need an $\epsilon$ very small already at energies of order $s_0\simeq 8$, the naive cutoff of the effective string theory. In this scenario, worldsheet particle production would be the dominant effect.

It would be very nice to extract the summand of the sum rule (\ref{sum_rule}) from the lattice results and see if the sort of toy model described here could have any resemblance with those, especially for the lower $N$ gauge theories.

\section{The Dual Blue Problem} \label{dualBlue}
To explore the space $(X,Y,Z)$ we used the radial parametrization  \cite{Monolith} -- that is we wrote 
\beq
(X,Y,Z)=r (\cos(\theta),\sin(\theta)\cos(\phi), \sin(\theta)\sin(\phi)) 
\eeq
and maximized $r$ at fixed angles -- and followed appendix D of \cite{dual2} closely; there the dual for a $Z_2$ problem with massive particles was written down so we should basically understand how to to take the $m\to 0$ limit there as well as adapt various normalizations to the bootstrap problem studied here. We did this. Since all steps are trivially adapted from \cite{dual2} we jump directly to the final result.

In the dual formulation, the maximum allowed radius $r$ is the result of a minimization problem so that 
\beq
r \le \int\limits_{0}^\infty ds\, \mathcal{L}[W_a] \label{dualR}
\eeq
for any functions $W_a$ normalized as  $\text{Im} (R_4)=0$ and
\beq
\cos\theta  \,R_1+\sin\theta\cos\phi  \,R_2+\sin\theta\sin\phi  \,\text{Re} (R_4) = 1
\eeq
with $R_a \equiv \pi \underset{s=i}{\text{Res}} W_a$. Except for $W_4$ the dual functions are automatically real in the imaginary axis; they can be parametrized as
\beq
W_{a} = \frac{1}{s^2+1} \sum_{m=0}^N (c^{(a)}_m +i \delta_{a=4} d^{(4)}_m) \Big(\frac{s-i}{s+i}\Big)^m
\label{dual_ansatz}
\eeq
with real constants $c$ and $d$. Importantly, contrary to the primal formulation, the dual bound (\ref{dualR}) is rigorous no matter what functions $W_a$ we plug there.  The larger $N$ is the better will the numerics be but in practice the converge is really good here; $N=4$ is plenty for all practical purposes. 

Finally, the dual functional $\mathcal{L}=\text{max} (A,B)$ with 
\beqa
\label{Afunc}
A&=&\frac{1}{\sqrt{2}} \sqrt{|W_3|^2+|W_{4'}|^2+|W_3-W_{4'}^2|}\\&&+\frac{1}{\sqrt{2}} \sqrt{|W_4|^2+2|W_{1}|^2+2|W_2|^2+|W_4^2-4W_1W_2|} \nn \\
B&=&\frac{1}{\sqrt{2}} \label{Bfunc} \sqrt{|W_3|^2+|W_{4'}|^2+|W_3-W_{4'}^2|}\\&&+\sqrt{|W_4|^2+|W_1-W_2|} \nn \,.
\eeqa
with $W_{4'}(s)=W_4(-s)$. 

The functional $B$ was overlooked in \cite{dual2} where $\mathcal{L}=A$ was used instead after extremizing over the $\lambda_a$ there; it turns out that in that case $A$ was always bigger than $B$ so that was luckily immaterial but a careful extremization over the lagrange multipliers $\lambda_a$ leads to $\text{max} (A,B)$ rather than just $A$. Indeed, for our massless blue problem we find that for fixed angles we might have $A>B$ for some energies and $A<B$ for some other energies. We need to live with a piece-wise defined function. 

The blue surface in figure \ref{MatrioskaFig} was generated using this dual method. 

\section{The Dual Red Problem}
\label{DualRed}

The dual formulation of the non-perturbative S-matrix bootstrap for weakly coupled theories at low energies was introduced in \cite{DualFT}, where the IR constraints were defined as point-wise conditions at the $s=0$ threshold as in \cite{FluxTube}. However, it turns out that it is not easy to generalize that formulation to bound the observables introduced in \eqref{XYZ}.

For this reason, we introduce and test different IR regulators for the dual EFT problem. In the next paragraphs, we will discuss a simple toy model:
the lower bound on $X$ without the inclusion of jet states.
This will serve as a playground to explain why the direct dualization of this problem is hard and why we need to introduce an IR regulator.

At the end of this section, we will apply one of these regulators to the full multi-particle problem and obtain the gray exclusion region in figure \ref{fig:dual_red_arcs}.

\subsection{Single component toy problem }

Consider the following toy problem. Take the massless $2\to2$ $S$-matrix element between fundamental branons, and ask what is the minimum of $S(i)$ with the EFT constraint that for $s\to 0$
\beq
S(s)=1+i \tfrac{s}{4}+\mathcal{O}(s^2).
\label{eq:pointwise_condition}
\eeq
The problem is to translate this constraint in terms of the values of the $S$-matrix on the real axis $s>0$, where it is bounded by unitarity.

\subsubsection{Point-wise IR constraints}

The condition \eqref{eq:pointwise_condition} is equivalent to the set of sum rules
\beqa
S(0)&=&S(i)+\frac{2}{\pi}\int\limits_0^\infty \frac{\Im S(z)}{z(1+z^2)}dz=1\nonumber\\
-i \partial_s S(s)|_{s=0}&=&-\frac{2}{\pi}\int\limits_0^\infty \frac{\Re S(z)-1}{z^2}dz=\frac{1}{4}\nonumber\\
\label{IR_sum_rules}
\eeqa

The Lagrangian for the problem can be compactly written as
\beqa
   \mathcal{L}&=&-S(i)\nonumber\\
    &+&\int\limits_0^\infty \Im (W(z) S(z))dz-S(i)\pi \underset{s=i}{\text{Res}} W \ \nonumber\\
    &+&\lambda_0(S(0)-1)+\lambda_1(-i \partial_s S(s)|_{s=0}-\tfrac{1}{4})\nonumber\\
    &+&\int\limits_0^\infty \mu(z)(1-|S(z)|^2)dz,
\label{dual_simple_lagrangian}
\eeqa
where the first line is the objective we maximize, the second line is the dispersive constraint, then we have the IR sum rules \eqref{IR_sum_rules}, and last line is the unitarity condition with $\mu>0$. $W$ is holomorphic in the upper half-plane and anti-crossing $W(-s^*)=-W^*(s)$, see \cite{Guerrieri:2020kcs}.

If we set $\pi\underset{s=i}{\text{Res}} W=(-1+\lambda_0)$, define the new variables 
\beqa
\Re \tilde{W}(s)&=&\Re W(s) + \frac{2\lambda_0}{\pi s(1+s^2)}, \nonumber\\
\Im \tilde{W}(s)&=&\Im W(s) -\frac{2\lambda_1}{\pi s^2},
\eeqa
and extremize over the primal variables, we get
\beq
-S(i)\leq -\lambda_0-\frac{\lambda_1}{4}+\int\limits_0^\infty \left(|\tilde{W}(s)|+\frac{\lambda_1}{2\pi s^2}\right)ds.
\label{dual_toy_pointwise}
\eeq
For generic values of $\lambda_0$ and $\lambda_1$, the functional diverges yielding the trivial bound $-S(i)\leq +\infty$.
To obtain stronger bounds, we need to expand \eqref{dual_toy_pointwise} around $s=0$ and impose the finiteness condition $\lambda_1\leq 0$. 

This problem is so simple that we have the luxury of guessing an analytic solution. We find $\lambda_0=16/81$, $\lambda_1=-64/81$ and
\beq
W=-\frac{130 i}{81 \pi (s^2+1)},
\eeq
which correctly gives $-S(i)\leq -\frac{7}{9}$ precisely matching (\ref{alphaBound}) with $\alpha=1/4$.

The issue with generalizing this method to the full problem \eqref{dualR} is to find the finiteness conditions for the dual functional in terms of the dual variables associated with the IR constraints. We were not able to find those conditions except for some special cases. For this reason, we have explored two different regulators that could be solved numerically.

\subsubsection{IR regulators: arc constraints}
\label{IR_regulator_arcs}

\begin{figure}[t!]
    \centering
    \includegraphics[width=0.45 \textwidth]{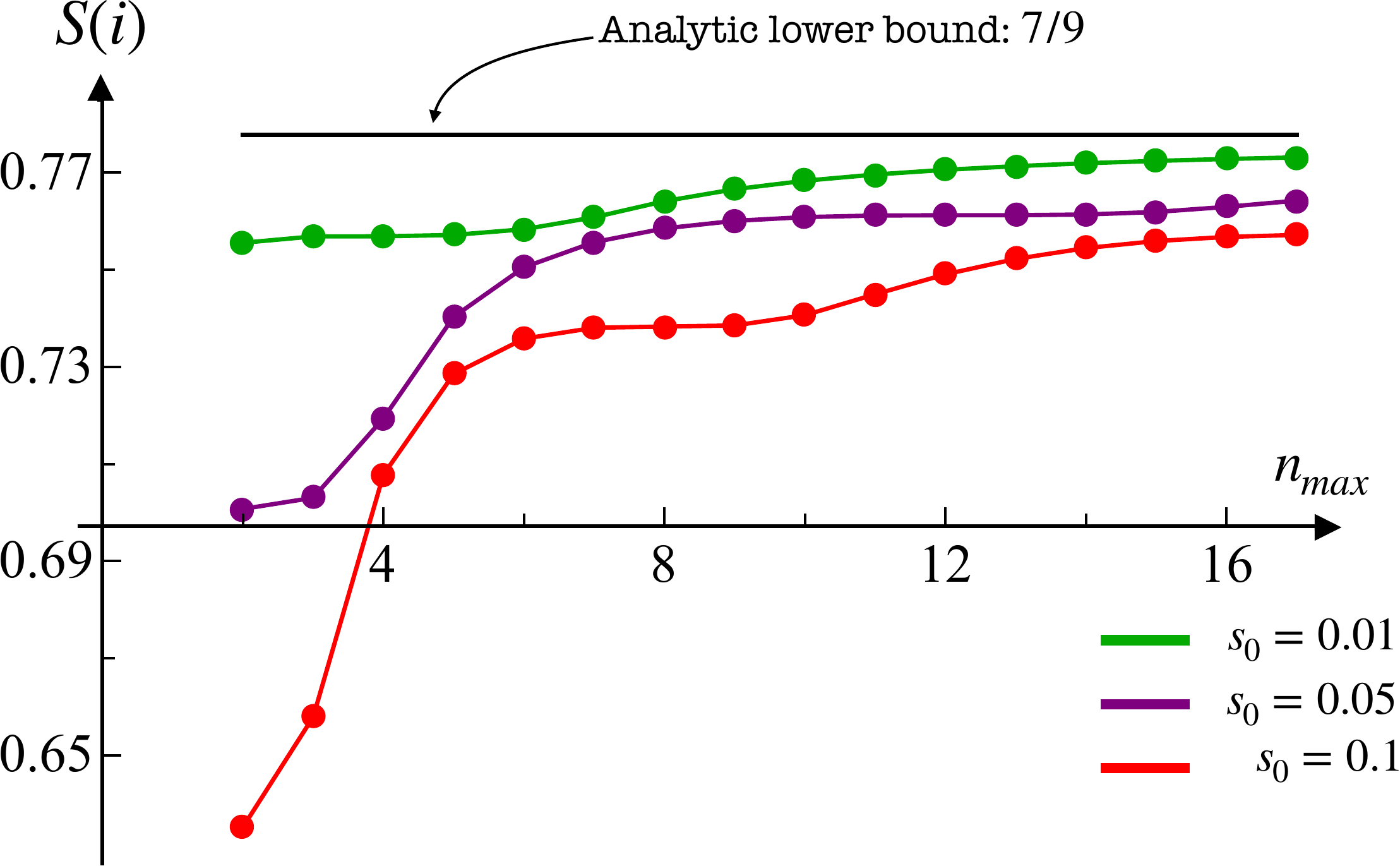}
    \caption{Bounds on the minimum of $S(i)$ for different values of the IR cutoff $s_0$, and as function of the number of dual parameters $n_\text{max}$.}
    \label{fig:plot_arcs_simple}
\end{figure}

The first regulator we test is based on the \emph{arc constraints} introduced in \cite{EliasMiro:2022xaa} for primal bootstrap problems. 
The idea is to replace \eqref{IR_sum_rules} with an approximate sum rule
\beq
\frac{1}{2}\int\limits_{s_0}^\infty \frac{\Im T(z)}{z^3}dz =\int\limits_\mathcal{A} dz \frac{T^{IR}(z)}{4\pi z^3}=\frac{\pi}{8}-\frac{s_0}{32},
\label{eq:arc_constraint}
\eeq
obtained by integrating $T/(2\pi i z^3)$ on the contour
\begin{figure}[h!]
    \centering
    \includegraphics[width=0.38 \textwidth]{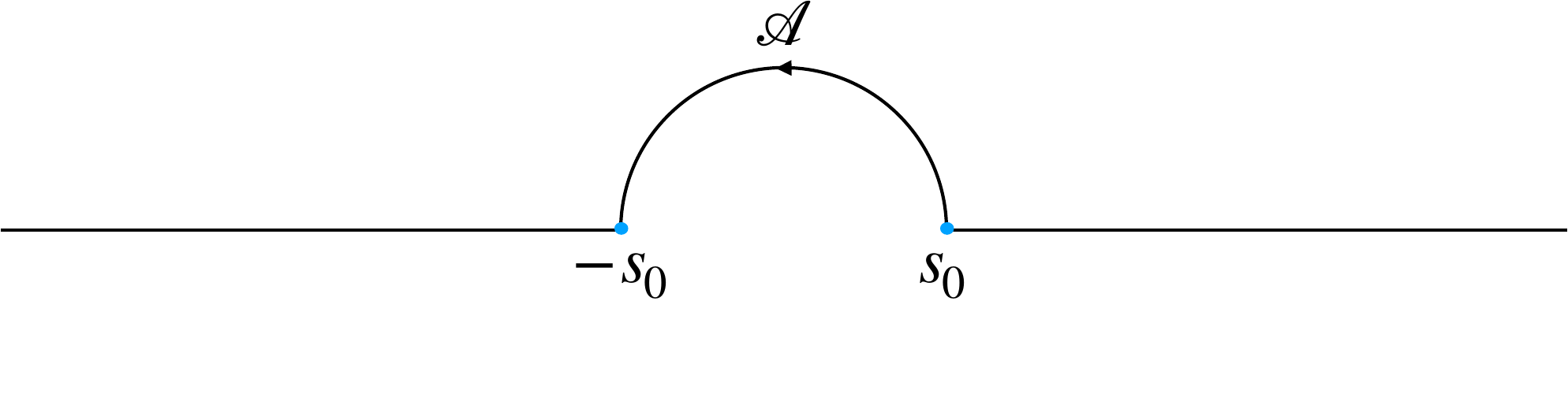}
    \label{fig:arc_plot}
\end{figure}
and replace the integrand over the arc $\mathcal{A}$ with 
\beq
T^{IR}(s)=\frac{s^2}{2}+i \frac{s^3}{16}.
\nonumber
\eeq
In terms of the S-matrix, the constraint \eqref{eq:arc_constraint} becomes
\beq
\frac{1}{2}\int\limits_{s_0}^\infty \frac{\Im T(z)}{z^3}dz=\frac{1}{s_0}-\int\limits_{s_0}^\infty \frac{\Re S(z)}{z^2}dz,
\label{eq:arc_constraint2}
\eeq
using the relation $T=-2 i s(S-1)$.
In the limit $s_0\to\infty$ the condition \eqref{eq:pointwise_condition} is equivalent to the point-wise constraints \eqref{eq:arc_constraint} and \eqref{eq:arc_constraint2}.

At this point, we just replace the point-wise derivative constraints with the arcs \eqref{eq:arc_constraint} and \eqref{eq:arc_constraint2}.
The dual problem then takes the form
\beq
-S(i)\leq \lim_{s_0\to 0} -\lambda_0\left( \frac{1}{ s_0}-\frac{\pi}{8}+\frac{s_0}{32}\right)+\int\limits_0^\infty |\tilde{W}(s)|ds,
\label{dual_func_arc}
\eeq
where
\beq
\tilde{W}(s)=W(s)+i\frac{\lambda_0}{s^2}\Theta(s-s_0), \quad \pi\underset{s=i}{\text{Res}} (W)=-1. \nonumber
\eeq

In figure \ref{fig:plot_arcs_simple} we numerically minimize the functional \eqref{dual_func_arc} for different values of the IR cutoff $s_0$, and as a function of $n_\text{max}$ that here stands for the number of terms in the dual ansatz -- see also \eqref{dual_ansatz}. The values of $S(i)$ below the curves are excluded for each fixed choice of the cutoff $s_0$. When both $s_0\to 0$ and $n_\text{max}\gg 1$ the dual exclusion bounds approach the analytic lower bound. A nice feature of this regularization is that for low $n_\text{max}$ and small $s_0$ we already have a good approximation of the optimal lower bound on $S(i)$.

\subsubsection{IR regulators: extended conditions}

\begin{figure}[t]
    \centering
    \includegraphics[width=0.45 \textwidth]{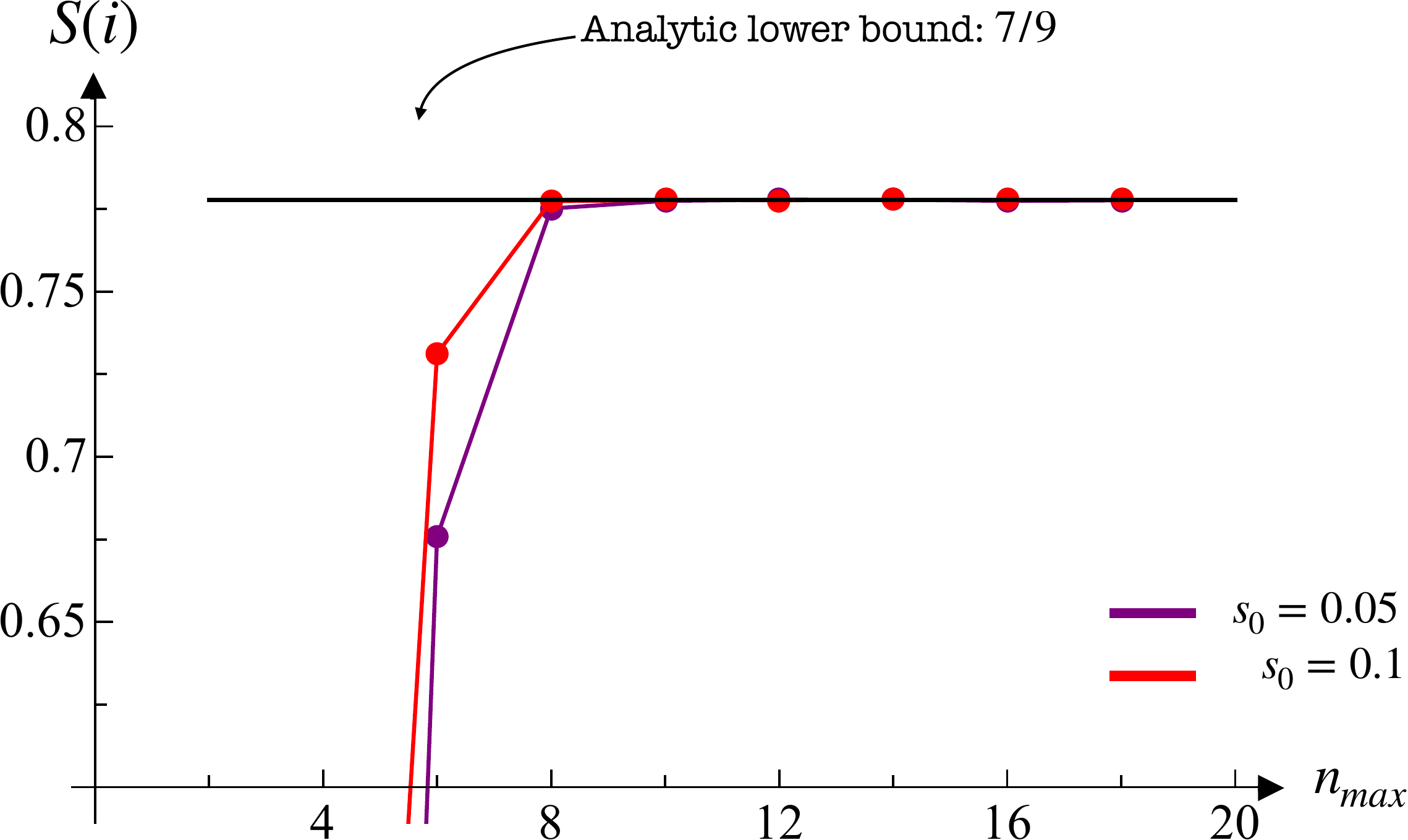}
    \caption{Bounds on the minimum of $S(i)$ for two different cutoffs $s_0$, and as a function of $n_\text{max}$, the total number of dual free variables in the ansatz. For smaller cutoffs convergence in $n_\text{max}$ becomes more difficult.}
    \label{fig:plot_extended_simple}
\end{figure}

The second IR regulator we tested was already introduced in the dual problem for four-dimensional scattering in \cite{EliasMiro:2023fqi}. The idea in this case is to replace
\beq
\Im S(s):= \Im S^{EFT}(s)= \frac{s}{4}, \quad s<s_0.
\label{extended_constraint}
\eeq
This strong condition is feasible since we always have the freedom of choosing the real or imaginary part of the S-matrix as an input of the dispersion relations, provided that $|\Im S|\leq 1$.

To get the dual, we plug the condition \eqref{extended_constraint} into the Lagrangian \eqref{dual_simple_lagrangian} and remove the point-wise constraints altogether obtaining
\beqa
-S(i)&\leq& \lim_{s_0\to 0} \int\limits_{s_0}^\infty |W(s)|ds +\label{extened_IR_functional}\\
&+& \int\limits_0^{s_0} \left(\frac{s}{4}\Re W(s){+}\sqrt{1{-}\frac{s^2}{16}}\Im W(s)\right)ds,\nonumber
\eeqa
where $\pi\underset{s=i}{\text{Res}} (W)=-1$. We have also used unitarity to replace $\Re S=\sqrt{1-s^2/16}$ for $s<s_0$.

\interfootnotelinepenalty=10000

Minimizing numerically the functional \eqref{extened_IR_functional} is harder than in the previous example. Since we have two different integration regions separated by $s_0$ it is better to define a mixed ansatz for $W$: one sum as in  \eqref{dual_ansatz} plus a contribution of the form 
\beq
W^\prime(s)=\frac{1}{s^2+1}\sum_{m=1}^{M} b_m(\rho_{s_0}^m(s)-\rho_{s_0}^m(-s)),
\eeq
where
\beq
\rho_{s_0}(s)=\frac{\sqrt{s_0}-\sqrt{s_0-s}}{\sqrt{s_0}+\sqrt{s_0-s}}.
\eeq
This sum is anti-crossing and analytic in the upper half plane, therefore compatible with the definition of $W$.

The numerical results are in figure \ref{fig:plot_extended_simple}. 
The dual bounds we obtain with this regulator are more accurate and for large $n_\text{max}$ the convergence looks perfect. 
This is expected since at fixed $s_0$ this constraint is stronger in imposing the EFT behavior. 
However, for small $n_\text{max}$ the bounds are weaker than those obtained using the arc constraints with the same number of variables. 

\subsection{The full dual problem with IR constraints}
\label{full_arc_problem}
It is time to derive dual bounds on the red rock inside the Matrioska. 
Here, we choose the arc regulator as it gives reasonable dual bounds with a small number of parameters $n_\text{max}$. 

First, we parametrize the $X,Z$ plane using radial coordinates
\beq
X=R_0+R\cos\theta,\quad Z=R\sin\theta.
\eeq
We need to shift $X$ because the red rock is not centered around the origin of the $X$ axis, but instead $7/9\leq X \leq 1$. We choose $R_0=8/9$. For fixed $\theta$ we optimize $R$.

Combining the results in \ref{dualBlue} with the arc constraints in \eqref{eq:arc_constraint} we can show that for fixed $\theta$ \footnote{For the off-diagonal amplitudes, the arc constraint is simply a null constraint of the form $\int\limits_{s_0}^\infty \Im T/z^3 dz=0$.}
\beqa
&&R\leq \lim_{s_0\to 0} \label{dualRedArcs}\\
&&-\sum_{a=1}^3 \lambda_{0,a}\left(\frac{1}{s_0}{-}\frac{\pi}{8}{+}\frac{s_0}{32}\right)+\max (A,B)-\pi R_0 \underset{s=i}{\text{Res}} (W_1).\nonumber
\eeqa
$A,B$ are given in \eqref{Afunc} and \eqref{Bfunc}, where we should replace $W_a$ with
\beqa
&&\tilde{W}_a(s)=W_a(s)+\lambda_{0,a}\frac{i}{ s^2}\Theta(s-s_0),\quad a=1,2,3,\nn\\
&&\tilde{W}_4(s)+W_4(s)=\frac{i}{2s^2}(\lambda_{0,4}{+}i\lambda_{0,5})\Theta(s-s_0)
\eeqa
and impose the linear constraints
\beqa
&& 1-\pi \cos\theta \underset{s=i}{\text{Res}} (W_1)-2 \pi \sin\theta \Re(\underset{s=i}{\text{Res}} (W_4)),\nn \\
&& \underset{s=i}{\text{Res}} (W_a)=0, \quad a=2,3,\nn \\
&& \Im(\underset{s=i}{\text{Res}} (W_4))=0.
\eeqa

Minimizing efficiently the functional \eqref{dualRedArcs} is a non-trivial task. 
The dual function is convex by construction. It is then sufficient to look for a local minimum. However it is non-smooth, and this usually causes the slowing down of gradient methods.

\begin{figure}[h!]
\includegraphics[width=0.45 \textwidth]{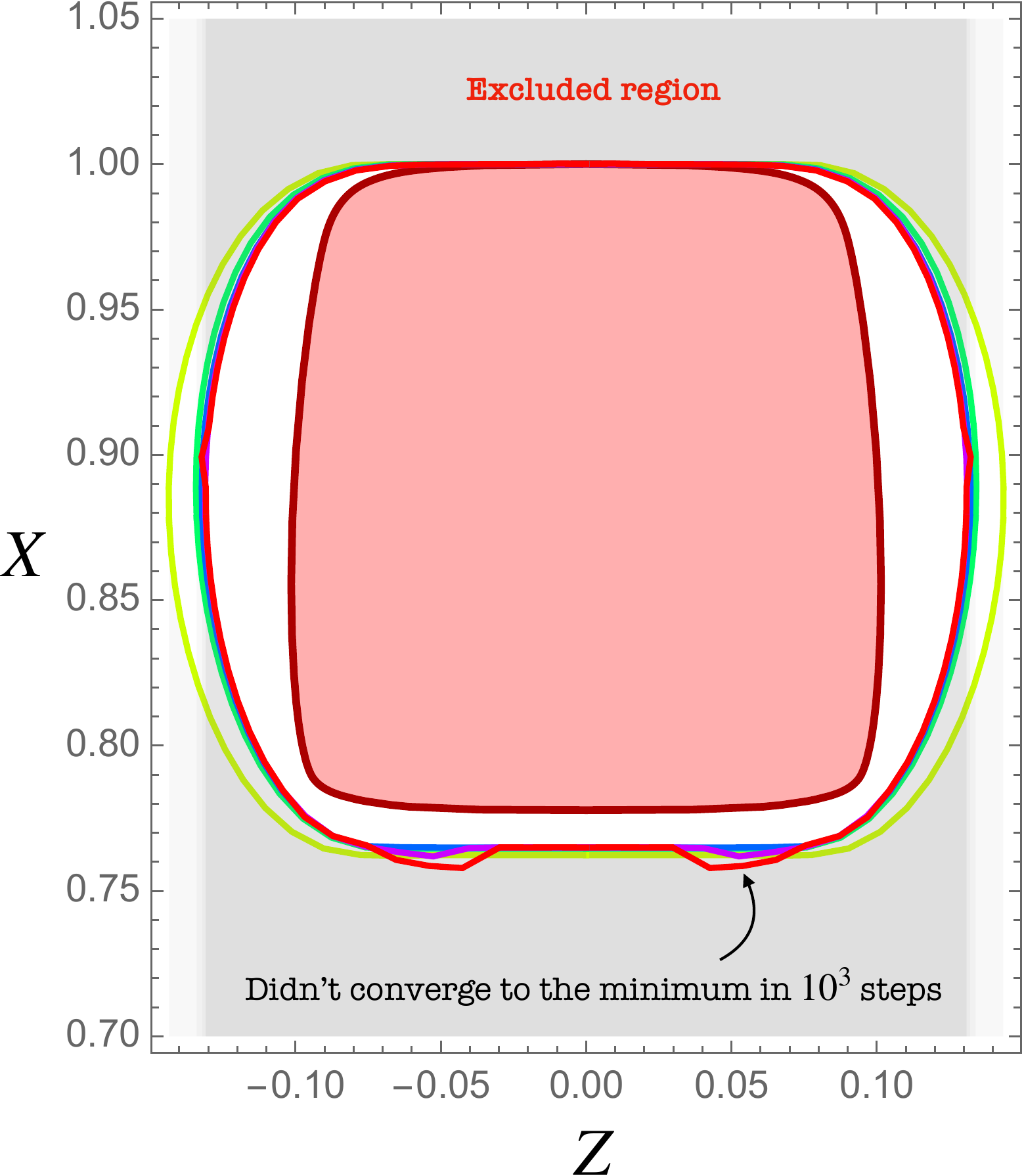}
    \caption{Exclusion bounds in the $X,Z$ space. The gray region is excluded, and the colored lines indicate the boundary of the excluded region for different values of $N$. The non-smoothness at the bottom is related to the slow convergence of the gradient algorithm used. We have minimized the functional \eqref{dualRedArcs} using the BFGS Quasi-Newton method with a cutoff on the maximum number of steps $10^3$. }
    \label{fig:dual_red_arcs}
\end{figure}

In figure \ref{fig:dual_red_arcs} we show the dual exclusion bounds obtained with arc constraints with $s_0=5\times 10^{-2}$. 
The gray region is ruled out, and the red region is ruled in using primal (which converged perfectly and corresponds to the true result of the physical question being asked in this letter). The different colored lines correspond to different values of $N$ in the ansatz \eqref{dual_ansatz} ranging from $N=1$ (light green) to $N=5$ (red). The gap in the bottom is due to numerical convergence (the minimum of $X$ was studied in the toy problem \ref{IR_regulator_arcs} where we can see a comparable gap for the same value of $N$). On the right, part of the gap might be due to the unitarity non-saturation in the primal discussed in section \ref{shiftsBoundary}. However, since dual numerics is still not optimal we cannot confirm this conjecture.

\section{Single component bounds for the Matrioska}
\label{analyticalbounds}

Coarse features of figure \ref{MatrioskaFig} can be derived analytically. For example, we have 

\begin{align*}
-1<{\color{blue}{X,Y,Z}}<1, &\\
7/9<{\color{red}{X,Y}}<1, &\\
7/9<{\color{green!60!black}{X}}< 1-\frac{48}{\gamma+217}, &\\
7/9<{\color{green!60!black}{Y}}<1-\frac{48}{\gamma/4+217}, &
\end{align*}
where we use a straightforward color code to refer to bounds applying to each of the solids in figure \ref{MatrioskaFig}. 

The first bound is of course trivial: it is just the Maximum Modulus Principle applied to the various single components. The {\color{red}{red}} bounds follow from the Schwarz-Pick lemma. To see how we recall the Schwarz-Pick inequality for functions from the UHP to the unit disk\footnote{Due to unitarity and polynomial boundedness, the S-matrix is such a function.} -- see also \cite{FluxTube}
\beq
\left|\frac{S(z)-S(w)}{1-S(z)\bar S(w)}\right| \leq \left|\frac{z-w}{z-\bar w}\right|
\eeq
In the limit $z=i \zeta \to 0$ and $w=i$ we get the inequality
\beq
-1-\frac{\zeta}{4}\frac{9S(i)-7}{S(i)-1}+\mathcal{O}(\zeta^2)\geq -1,
\eeq
from which $7/9<S(i)<1$ follows.

The {\color{green!60!black}{green}} bounds are also a consequence of the Schwarz-Pick lemma, albeit in a more non-trivial way: 
again following \cite{FluxTube}, we introduce an auxiliary map from the UHP to the unit disk $\chi=(\alpha i-s)/(\alpha i+s)$, and consider the function $S(s(\chi)) \equiv \mathbf{S}(\chi)$. 
Define the following functional
\beq
\Delta[\mathbf{S}](z|w)=\frac{\mathbf{S}(z)-\mathbf{S}(w)}{1-\mathbf{S}(z)\overline{\mathbf{S}(w)}}\bigg/\frac{z-w}{1-\bar w z},
\eeq
where $\mathbf{S}:\mathbb{D}\rightarrow\mathbb{D}$ is holomorphic. The Schwarz-Pick inequality then reads $|\Delta[\mathbf{S}](z|w)|\leq 1$. Note that $\Delta[\mathbf{S}](z|w)$ is also a holomorphic map on the disk, therefore it also obeys the Schwarz-Pick inequality, and we can iterate. For instance
\beq
\lim_{w\to 1}\lim_{z\to 1}|\Delta[\Delta[\mathbf{S}](z|\chi(i))](z|w)|\leq 1.
\eeq
Evaluating the limit, and using the EFT expansion $S(s)=e^{i s/4+i\gamma s^3}+\mathcal{O}(s^4)$, we obtain
\beqa
1&-&\frac{(z{-}1)(w{-}1)(S(i)+1)\alpha^2}{384(S(i)-1)(9S(i)-7)}\times \nonumber \\
&&(\gamma(S(i){-}1)+217 S(i){-}169)+\dots \leq 1.
\eeqa
which is the inequality quoted above.

For the value of $\gamma= 0.768$, chosen to derive the green matrioska, we obtain $\tfrac{7}{9} <S(i)< \tfrac{21221}{163\times 167}\simeq 0.7796$.

Note that these simple single component bounds manifest how the {\color{green!60!black}{green}} matrioska is $O(100)$ times smaller in linear dimension compared to the red surface. 

\phantom{That's all folks}

\end{document}